\documentclass[preprint,preprintnumbers,aps,showpacs,superscriptaddress,citeautoscript]{revtex4-1}

\usepackage[utf8]{inputenc}
\usepackage{amsmath}
\usepackage{graphicx}
\usepackage{amssymb}
\usepackage{times}
\usepackage{color}
\usepackage{float}
\usepackage{multirow}
\usepackage{subcaption}
\usepackage{braket}
\usepackage{hyperref}
\usepackage{acronym}

\begin{document}

\title{ADAQ: Automatic workflows for magneto-optical properties of point defects in semiconductors}

\author{Joel Davidsson}
\email{joel.davidsson@liu.se}
\affiliation{Department of Physics, Chemistry and Biology, Link\"oping University, Link\"oping, Sweden}
  
\author{Viktor Iv\'ady}
\affiliation{Department of Physics, Chemistry and Biology, Link\"oping
  University, Link\"oping, Sweden}
\affiliation{Wigner Research Centre for Physics, Budapest, Hungary}

\author{Rickard Armiento}
\affiliation{Department of Physics, Chemistry and Biology, Link\"oping University, Link\"oping, Sweden}

\author{Igor A. Abrikosov}
\affiliation{Department of Physics, Chemistry and Biology, Link\"oping University, Link\"oping, Sweden}

\begin{abstract}

Automatic Defect Analysis and Qualification (ADAQ) is a collection of automatic workflows developed for high-throughput simulations of magneto-optical properties of point defects in semiconductors. These workflows handle the vast number of defects by automating the processes to relax the unit cell of the host material, construct supercells, create point defect clusters, and execute calculations in both the electronic ground and excited states.
The main outputs are the magneto-optical properties which include zero-phonon lines, zero-field splitting, and hyperfine coupling parameters.
In addition, the formation energies are calculated.
We demonstrate the capability of ADAQ by performing a complete characterization of the silicon vacancy in silicon carbide in the polytype 4H (4H-SiC).
\end{abstract}

\maketitle


\section{Introduction}

Point defects in wide-bandgap semiconductors have spanned a wide range of applications, including but not limited to qubit realizations~ \cite{Childress:Science2006,Jelezko:PSS2006,Jacques2009,Hanson:Nature2008,Awschalom2013}, biosensors~\cite{mcguinness2011quantum,Kucsko2013,Balasubramanian:Nature2008}, 
accurate chemical sensors~\cite{aslam2017nanoscale}, nanoscale electric field and strain sensors~\cite{Falk2014}, and nano thermometers~\cite{anisimov2016optical}.
Most of these applications have been realized with the NV center in diamond~\cite{Davies:PRSLA1976,gruber1997scanning,dobrovitski2013quantum} -- a point defect cluster in diamond consisting of a carbon vacancy and a nitrogen substitution.
Recently, other point defects in diamond (such as the silicon vacancy cluster and the related group 14 vacancy clusters~\cite{hepp2014electronic,thiering2018ab}), as well as point defects in other materials (such as divacancy and silicon vacancy in silicon carbide (SiC)~\cite{methodology_paper,ivady2017identification,6H} and boron vacancy in boron nitride (BN)~\cite{gottscholl2020initialization,ivady2020ab}), have attracted remarkable interest.
Other not yet discovered point defects in various semiconductor hosts may have even more attractive properties for existing as well as novel applications.

Due to the vast number of possible point defects, to discover novel potentially interesting candidates is a challenging inverse design problem~\cite{bassett2019quantum}.
The latter means selecting the desired properties and letting the structure and materials vary.
A potential starting point is to use high-throughput calculations and collect the results in a database.
Previous high throughput works in this direction focused on the energetics of point defects~\cite{goyal2017computational,broberg2018pycdt}.
Furthermore, these high-throughput workflows handle only single defects, whereas defect clusters, such as pair defects, are among the most studied defects for quantum applications.

To find and identify point defects is a labor-intensive process.
Point defect identification in semiconductors may be achieved by comparing calculated magneto-optical properties with experimental values~\cite{methodology_paper}.
The considered magneto-optical properties may include zero phonon line (ZPL), hyperfine coupling parameters, and zero field splitting (ZFS).
For the ZPL, both the energy, polarization, and intensity of the line provide information about the workings of the point defect.
A point defect can exist in different configurations depending on the host material.
For reliable identifications, one should look at different configurations and different charge and spin states of each defect.
A suitable automatic workflow can take all this into account and fully characterize a point defect.
In addition, the formation energy can be calculated to characterize the stability of different point defects.
All these data can assist the researcher to better understand the point defect and evaluate if it is useful for a given application.

In this paper, we present the full characterization workflow, which is the primary component of ADAQ, that allows for high-throughput calculations of magneto-optical properties of point defects and their clusters of arbitrary size for finding and identifying potentially interesting systems.
Furthermore, ADAQ also contains a simplified workflow for quick estimates of the ZPL and energy of point defects intended for high-throughput screening of defects to identify candidates for which the full characterization workflow can be deployed.
The present work goes beyond similar prior efforts in its focus on magneto-optical properties and point defect clusters, both of which have not been considered earlier in the context of automated workflows.
We demonstrate how these properties are obtained with ADAQ considering a well-known defect: the silicon vacancy in 4H-SiC~\cite{ivady2017identification}.

The outline of the paper is the following.
Section~\ref{sec:theory} introduces the properties calculated in ADAQ and the theory behind the calculations.
Section~\ref{sec:method} describes the software used to set up and execute the calculations, as well as the settings for the first principle software used.
In addition, an overview of ADAQ and its default input parameters, such as the size of the supercell, are presented, as well as details for the unit cell and host workflows.
The full characterization workflow is described in Section~\ref{sec:workflow}.
Here, overviews are presented for both the ground and excited state workflows as well as details about each step in the workflows.
The workflow results are stored in a database, and Section~\ref{sec:database} outlines which properties are stored and how.
Section~\ref{sec:resuls} shows the information that ADAQ produces for a point defect, illustrated by the silicon vacancy in 4H-SiC.
The strengths and limitations of the full characterization workflow are examined in Section~\ref{sec:discussion} and conclusions are presented in Section~\ref{sec:conclusion}.
Appendix~\ref{appendix:defect_gen} shows the algorithm for defect generation which constructs point defect clusters up to an arbitrary size.
Appendix~\ref{appendix:screen} outlines the screening workflow that produces quick estimates of the ZPL and energy of the point defect.
There is also a list of acronyms at the end of the paper.

\section{Theory}
\label{sec:theory}

The subsections below present the theoretical background on first-principles calculations of the point defect properties built-in into ADAQ.

\subsection{Photoluminescence}

\begin{figure*}[h!]
   \includegraphics[width=\textwidth]{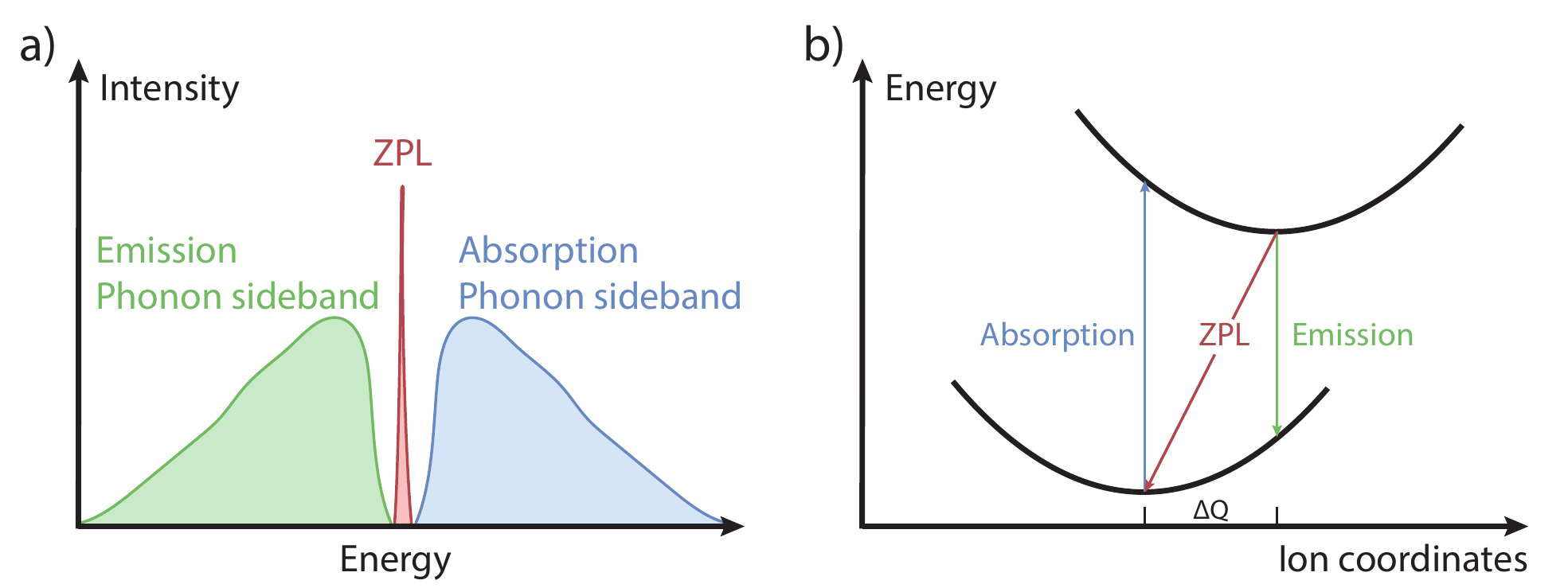}
	\caption{a) Schematic picture of a photoluminescence spectrum. The ZPL is a sharp line between the two broader sidebands. b) Excitation cycle which produces the ZPL. The lower (higher) parabola represents the ground (excited) state energy surface. The displacement between different ion minima in the ground and excited state is $\Delta Q$.}
	\label{fig:pl} 
\end{figure*}

Point defects in semiconductors may introduce states in the band gap.
If an optical transition between these states is allowed and the non-radiative decay rate is low, a ZPL will appear in the photoluminescence spectrum.
One of the main tasks of ADAQ is to calculate whether a ZPL exists for a given defect and predict its energy.
Figure~\ref{fig:pl} a) and b) shows a schematic photoluminescence spectrum for a defect at low temperature and the different transitions between two defect states, respectively.

\subsubsection{ZPL}

The sharp peaks in the photoluminescence spectra, like the one in Figure~\ref{fig:pl}a), are called ZPL, which arise from the direct transition between the defect states in the band gap.
They can be used to identify the point defect present in the material.
As seen in Figure~\ref{fig:pl}b), the excitation cycle starts with the absorption of a photon that excites the system from the ground state to the excited state, followed by a relaxation of the ions to the excited state equilibrium configuration.
In the excited state, there are two possibilities for radiative decay back to the ground state.
The state can either decay straight back to the ground state minimum with no phonon contribution, which produces the ZPL, or decay to a higher energy atomic configuration and relax to the lowest energy configuration through phonon emission.
Given the energy for the ground and excited states, the ZPL energy is defined as:
\begin{equation}
E_{ZPL}=E_{e,min}-E_{g,min}
\end{equation}
where $E_{e,min}$ ($E_{g,min}$) is the excited (ground) state in its corresponding minima.
Here, the two states are assumed to have similar phonon spectra, and the zero-point energy of the phonon modes cancels out between the ground and excited states.
As mentioned before, the ZPL exists as long as the non-radiative decay is small.
The non-radiative effects are not taken into account in the calculations.
The rate of non-radiatively relaxation increases exponentially as the energy difference between excited and ground states decreases.
Hence, predicted ZPLs below 0.6 eV are considered to be uncertain.

If a ZPL exists, additional factors could be used to determine if the defect is a promising candidate for a particular application, such as a single photon emitter.
Three additional properties that can be used for the characterization include the transition dipole moment (TDM) that describes the polarization and intensity of the ZPL~\cite{davidsson2020theoretical}, the ion relaxation between the ground and excited state which indicates the ratio between the ZPL and the phonon sideband emission, and the spontaneous macroscopic polarization which tells if the ZPL is stable against electric fields.

\subsubsection{Intensity of ZPL}

The TDM is calculated between the single particle orbitals of the transition and is defined as:
\begin{equation}
\boldsymbol{\mu}=\braket{\psi_f|q\mathbf{r}|\psi_i}=\frac{i \hbar}{(\epsilon_{f,k}-\epsilon_{i,k})m}\braket{\psi_{f,k}|\mathbf{p}|\psi_{i,k}}.
\end{equation}
Here, $q\mathbf{r}$ is the dipole operator between the initial $\psi_i$ and final state $\psi_f$, which can be rewritten into reciprocal space with the momentum operator $\mathbf{p}$ instead and a prefactor with the eigenvalue difference between the final and initial states, $\epsilon_{f,k} - \epsilon_{i,k}$.
The constants used in this formula are the Planck constant $\hbar$ and the mass of the electron $m$.
The $\boldsymbol{\mu}$ is calculated for all optical transitions in the excitation cycle and from it, the optical polarization of absorption, ZPL, and emission are extracted as well as the intensity $|\bar{\boldsymbol{\mu}}|^2$.
The optical lifetime $\tau$ is calculated from the intensity $|\bar{\boldsymbol{\mu}}|^2$~\cite{davidsson2020theoretical} with the following equation 
\begin{equation}
\tau=\frac{3 \epsilon_0 h c^3}{n 2^4 \pi^3 \nu^3 |\bar{\boldsymbol{\mu}}|^2},
\label{eq:lifetime}
\end{equation}
where the $\epsilon_0$ and $c$ are the vacuum permittivity and the speed of light, respectively, $\nu$ is the transition frequency for the specific transition in question (absorption, ZPL, or emission) and $n$ is the refractive index, which for 4H-SiC is $2.6473$.
This method has been applied to the divacancy in 4H-SiC, where the inclusion of the ion relaxation and the corresponding electronic change of the excited state is crucial to include when calculating $\boldsymbol{\mu}$ for the ZPL~\cite{davidsson2020theoretical}.

\subsubsection{Ion relaxation}

An additional desired property for a bright single photon emitter is a large ratio between the photon count of the ZPL and the phonon emission sideband.
This ratio can be determined from the Huang-Rhys factor, which calculates the coupling between the electronic and vibronic states~\cite{huang2000theory}.
This factor is expensive to calculate.
Hence a 1D model with two measures of ion relaxation between the ground and excited state has been introduced~\cite{alkauskas2014first} and tested~\cite{tawfik2017first}.
These two measures are the squared displacement of the ions $\Delta R$, and a parameter in which each displacement is scaled with the atom weight $\Delta Q$, which is shown in Figure~\ref{fig:pl}b).
$\Delta R$ is calculated as:
\begin{equation}
(\Delta R)^2 = \sum_{a}{|\mathbf{\bar{R}}_{ea}-\mathbf{\bar{R}}_{ga}|^2},
\end{equation}
where $\mathbf{\bar{R}}_{ea}$ and $\mathbf{\bar{R}}_{ga}$ are the ion positions in the excited and ground state, respectively.
The summation runs over all ions in the supercell.
$\Delta Q$ is calculated as:
\begin{equation}
(\Delta Q)^2 = \sum_{a}{m_a}{|\mathbf{\bar{R}}_{ea}-\mathbf{\bar{R}}_{ga}|^2},
\label{eq:dq}
\end{equation}
where $m_a$ is the atomic mass.

\subsubsection{Spontaneous macroscopic polarization}

For an ideal single photon emitter, the ZPL should be stable against electric fields, \emph{i.e.}, show no spectral diffusion~\cite{udvarhelyi2019spectrally}.
This can be achieved if the spontaneous macroscopic polarization, that couples the defect to external electric field fluctuations, is small.
The spontaneous macroscopic polarization is the difference between the ground and excited state macroscopic polarization which is calculated from the Berry phase according to the modern theory of polarization~\cite{king1993theory,resta1992theory,spaldin2012beginner}.
However, one may need to do additional calculations to ensure that no wrap-around error is present.

\subsubsection{Defect ionization and affinity energy}
\label{sec:iae}

Depending on the positions of the defect states in the band gap with respect to the conduction and valence band edges, it is possible that an optical excitation can change the charge state of the point defect.
This can happen either if an electron moves from a defect state to the conduction band edge (bound-to-free transition), \emph{i.e.} defect ionization energy, or if an electron moves from the valence band edge to a defect state (free-to-bound transition), \emph{i.e.} defect affinity energy.
If any of these transitions require lower energy than the ZPL, the point defect will change the charge state before emitting a ZPL.

\subsection{Hyperfine coupling parameters and ZFS}

For point defects with spin (unpaired electrons), additional interactions occur.
The unpaired electrons, which have magnetic moments, interact with magnetic fields and other magnetic moments.
If some ions in the material have spin (a non-zero nuclear g-factor, which are listed at easyspin website~\cite{easyspin} for all isotopes), the magnetic moments of the ion and electron can couple and give rise to hyperfine interaction.
The hyperfine interaction~\cite{Szasz2013} is defined as:
\begin{equation}
H_{hyperfine}=\mathbf{\hat{S}^TA\hat{I}}.
\end{equation}
Here $\mathbf{\hat{S}}$ is the electron spin operator, $\mathbf{\hat{I}}$ is the nuclear spin operator, and $\mathbf{A}$ is the hyperfine tensor.
We define $\mathrm{A_{xx}}$, $\mathrm{A_{yy}}$, and $\mathrm{A_{zz}}$ as the diagonal components of the $\mathrm{A}$ tensor and $\mathrm{A_{z}}$, the projection on the $z$-axis, as the hyperfine splitting.
The tensor only exists if both the spin of the defect and ions are non-zero.
In practice, this tensor is approximated by calculating the Fermi-contact interaction, which depends on the spin density at the center of the nucleus, and the dipole-dipole interaction.

If the point defect has more than one unpaired electron (at least spin-1), the unpaired electrons interact with each other and separate states with different absolute spin quantum numbers without a magnetic field present, producing a ZFS.
The ZFS comes from the D-tensor.
The D-tensor describes the interaction for the total effective spin, which can be approximated with spin-spin dipole interaction in semiconductors of light elements~\cite{Ivady2014} and is defined as:
\begin{equation}
H_{ZFS}=\mathbf{\hat{S}^TD\hat{S}}.
\end{equation}
Here $\mathbf{\hat{S}}$ is the electron spin operator and $\mathbf{D}$ is the D-tensor which is traceless and symmetric.
Usually, the diagonal elements of the D-tensor are ordered in increasing order, where the $z$ component is the largest.
From the D-tensor, the ZFS is calculated as $\frac{3}{2}D_z$ for spin-1 defects.
A detailed description of the methodology for the D-tensor calculation can be found in Ref.~\onlinecite{Ivady2014}.

\subsection{Formation energy and charge-state transition levels}

The formation energy is the energy needed to create the defect.
It can be useful when comparing different defects to see which one is the most stable.
The charge-state transition levels tell us where the Fermi energy needs to be in the material in order to keep the defect in a given charge state. 
The formation energy for charged defects is defined as~\cite{FreysoldtRMP2014,goyal2017computational}:
\begin{equation}
\Delta H_{D,q}(E_f,\mu) = [E_{D,q}-E_H] + \sum_i{n_i\mu_i} + qE_f + E_{\mathrm{corr}}(q).
\label{eq:form}
\end{equation}
The formation energy $\Delta H_{D,q}(E_f,\mu)$ depends on the Fermi energy $E_f$ and the chemical potential $\mu$, both is discussed further below.
$E_{D,q}$ and $E_H$ are the total energies for the charged defected supercell and the perfect host supercell, respectively.
The variable $n_i$ keeps track of the atoms of chemical element $i$ added ($n_i<0$) or removed ($n_i>0$), $q$ is the charge of the defect supercell and $E_{\mathrm{corr}}$ is the correction term needed to minimize finite-size effects.

From the formation energy, charge-state transition levels can be found~\cite{FreysoldtRMP2014}.
These levels show where a transition from one charge state to another occurs and are defined as:
\begin{equation}
\epsilon(q_1,q_2) = \frac{\Delta H_{D,q1}-\Delta H_{D,q2}}{q_2-q_1} = \frac{E_{D,q1}+E_{corr}(q1)-E_{D,q2}-E_{corr}(q2)}{q_2-q_1},
\end{equation}
where $E_f$ is set to zero.

\subsubsection{Chemical potential}

In Eq.~\eqref{eq:form}, the chemical potentials for the elements involved are needed.
The chemical potentials used in this paper are calculated with the same settings as in the workflow for all elements.
This calculation gives the total energy (calculated at zero temperature) for each element in the periodic table.
The structures for the elements are taken from Ref.~\onlinecite{lejaeghere2014error}, which has a list of all the elements in their periodic ground state structure at zero temperature.
Therefore, all elements are treated equally, and thus the reference states for e.g., N2 and O2 are not isolated molecules, but the 0K elemental structures (which for these are dimers in a periodic structure).
Furthermore, we use the energies as calculated by VASP without molecular energy correction, which is often used in similar phase diagrams (see e.g., Ref.~\onlinecite{wang2006oxidation}).
The total energy will give an upper limit to the chemical potential.
For SiC, the silicon and carbon ground state will give an upper limit to the chemical potential, denoted a rich phase ($\mu^{\mathrm{rich}}$).
In addition, the following relation holds: $\mu_{\mathrm{SiC}}=\mu_{\mathrm{Si}}+\mu_{\mathrm{C}}$.
Using the upper limit for one chemical element, the lower limit for the other chemical element can be obtained from this relation.
Here, one assumes that there is only one stable phase (SiC) between the elements (Si, C) on the convex hull.
If there would be other stable phases on the convex hull, like SiC$_2$, one should consider these like in Ref.~\onlinecite{broberg2018pycdt}.

\subsubsection{Potential alignment}

The formation energy is plotted against the Fermi energy, and most often, the valence band maximum (VBM) is set to zero.
This convention means that the endpoint of the Fermi energy is the bandgap energy, \emph{i.e.}, conduction band minimum (CBm).
However, when comparing the host supercell to a charged defect supercell, the VBM and CBm may be shifted.
This shift can be accounted for by comparing the average potential far away from the defect.
In this work, the supercells are large enough that potential alignment is negligible.
Thus the VBM and CBm are taken straight from the host supercell.

\subsubsection{Charge correction}

When comparing the energy of periodic cells of different charge states, one has to account for  the self-interaction energy contribution of the extra charge.
Different charge correction schemes have been suggested: Makov-Payne (MP)~\cite{makov1995periodic}, Lany-Zunger (LZ)~\cite{LanyZunger08}, and  Freysoldt–Neugebauer–Van de Walle (FNV)~\cite{Freysoldt}.
FNV is the most accurate but non-trivial to use for defect clusters and might run into computational difficulties~\cite{komsa2012finite}, making this correction challenging to use in high-throughput frameworks.
Hence, we choose the LZ correction scheme that gives the same correction for all defects.
The LZ correction $E_{corr}$ is defined as
\begin{equation}
E_{\mathrm{corr}} = (1+f)\frac{q^2 \alpha_M}{2 \boldsymbol{\epsilon} L},
\label{eq:cc}
\end{equation}
where $f$ is a proportionality factor linking the third-order correction to the first-order, $q$ is the charge, $\alpha_M$ is the Madelung constant, $\boldsymbol{\epsilon}=4 \pi \epsilon_0 \epsilon_r$ where $\epsilon_r$ is the dielectric constant, and $L$ is the length of the supercell ($L=V^{1/3}$).

\section{Methodology}
\label{sec:method}

The High-Throughput Toolkit (\textit{httk})~\cite{httk,Armiento2020} is used for automatic control of the ADAQ calculations.
It is a framework that handles transferring calculations between a local computer and a supercomputer and executing the runs by running taskmanagers.
The taskmanager includes checkers that monitor the runs, ensure that the runs converge as intended, and cancel any run that breaks the predefined rules.
Any software can be controlled by \textit{httk}, but for ADAQ, all workflows are presently implemented only for the Vienna Ab initio Simulation Package (VASP)~\cite{VASP,VASP2}.

VASP runs density functional theory (DFT)~\cite{Hohenberg64,Kohn65,ivady2018first} calculations with projector augmented wave (PAW)~\cite{PAW,Kresse99} method.
For the excited state calculation, the constrained occupation DFT method~\cite{cDFT,Gali:PRL2009} is used.
This method is also known as $\Delta$SCF~\cite{cDFT,SCFrev,SCFsum} when using the total energy difference between the ground and excited state.
Formally, one should use the DFT formalism of Görling (generalized adiabatic connection\cite{GACDFT}) to include stationary densities and handle the excited states~\cite{davidsson2021color}.
However, this formalism introduces an orbital-dependent exchange-correlation functional.
In practical calculations, this functional is approximated by the exchange-correlation functional approximation from the standard DFT formalism.
The exchange-correlation effects are described by the semi-local functional of Perdew, Burke, and Ernzerhof (PBE)~\cite{PBE}.
For calculating ZPL, we use the method described in detail and tested for the divacancy point defect in 4H-SiC in Ref.~\onlinecite{methodology_paper}.
In this previous paper, we concluded that using the PBE functional for a 576 atom supercell with $2 \times 2 \times 2$ k-point set provides a good compromise between the accuracy and efficiency, which is suitable for high-throughput calculations.
Due to the use of the PBE functional, we have found a systematic underestimation of about 0.2 eV for the calculated ZPL compared with experiment for 4H-SiC.
The same offset can be seen for the neutral divacancy and charged silicon vacancy in 6H-SiC~\cite{6H}.
The hybrid functional of Heyd, Scuseria, and Ernzerhof (HSE06)~\cite{HSE03,HSE06} corrects this error but is not suitable for high-throughput calculations at present due to its high computational cost~\cite{methodology_paper}.
To compare the ADAQ results with experimental data, one should add this systematic shift of 0.2 eV to the estimated ZPL.
If higher accuracy is needed, HSE calculations can be run on top of the PBE results which produces accurate results as tested in Ref.~\onlinecite{methodology_paper}.
For even greater accuracy, fully self-constituent HSE or other higher-order methods, such as GW, are possible.
Workflows using these higher-order methods are presently not part of ADAQ, however, such an extension would be relatively straightforward.

When running high-throughput calculations for a wide range of different elements, convergence settings must be chosen to match the requirements of the most demanding chemical elements.
To ensure sufficient convergence for all elements, the plane wave energy cutoff is set to 600 eV and kinetic energy cutoff to 900 eV.
These values are in the middle of the range recommended for the more demanding PAW pseudopotential and are slightly larger than those used in the Materials Project (520 eV)~\cite{jain2011high,jain2011formation}.
Unless specified otherwise, we apply the following settings:
\begin{itemize}
\item The Fast Fourier transform (FFT) grid is set to twice the largest wave vector.
\item The k-point grid is constructed by $\Gamma$ centered Monkhorst-Pack~\cite{monkhorst1976special}.
\item When only $\Gamma$-point is used, Fermi smearing with a width of 1 meV, otherwise the tetrahedron method with Blöchl corrections~\cite{blochl1994improved} is used.
\item For defect calculations, symmetry is not used.
\item Non-spherical contributions of the PAW spheres are included.
\item Projection is done in real space with automatic optimization.
\item The ions are updated using the quasi-Newton method.
\end{itemize}

\begin{figure*}[h!]
   \includegraphics[width=\textwidth]{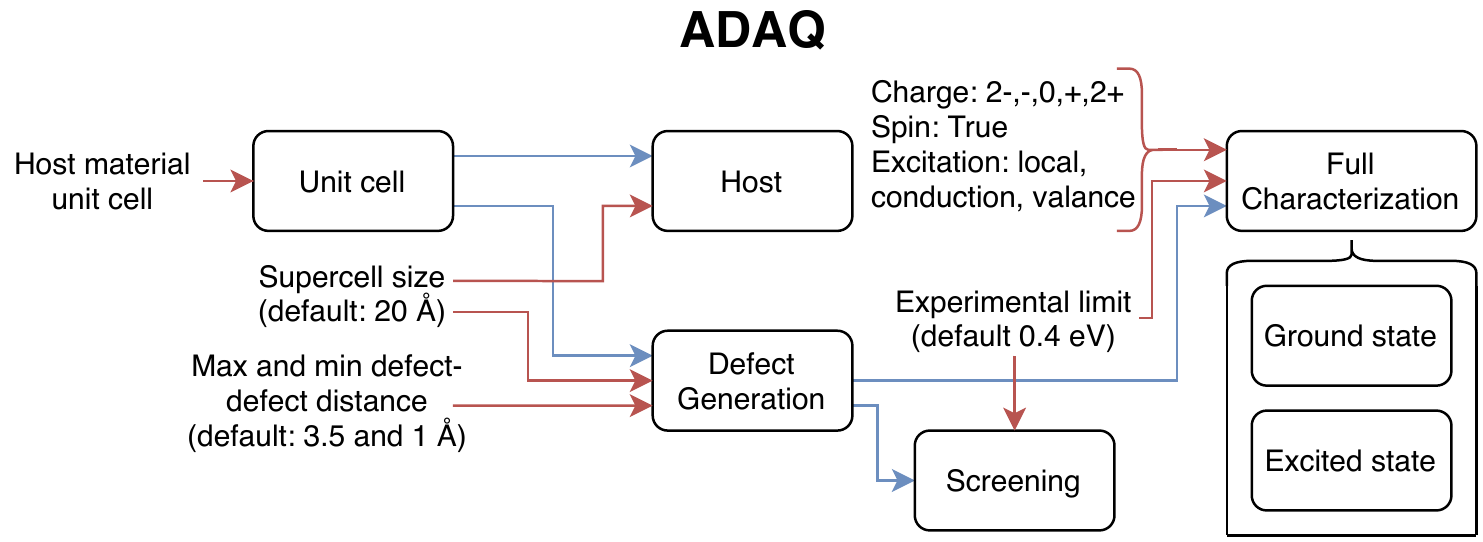}
	\caption{Overview of ADAQ, where the workflows are represented with boxes. Red arrows represent user input to the workflows. The blue arrows represent the output from one workflow to the next workflow. The host and defect generation both need the relaxed unit cell, and the screening and full characterization both need the point defect supercells. The full characterization workflow consists of two workflows, the ground and excited state workflows.}
	\label{fig:ADAQ} 
\end{figure*}

Using \textit{httk} and VASP, the workflows are constructed for ADAQ.
Figure~\ref{fig:ADAQ} shows an overview of all the workflows included in ADAQ with their corresponding inputs.
Computational details for the two smaller workflows, the unit cell and host workflows, are presented below.
Details of the more extensive workflows, ground and excited state workflows, are presented later in this paper.

Here are the details for the two smaller workflows in ADAQ, the unit cell and host.
First, the unit cell of the chosen host material is relaxed.
The unit cell workflow carries out volume relaxations with PBE functional as default.
To ensure accurate volume, these calculations are executed with a k-point set $10 \times 10 \times 10$, Gaussian smearing with a width of 1 meV is used, the plane wave energy cutoff is set to 700 eV, the plane wave kinetic energy cutoff is 1400 eV, the electronic convergence criterion is $10^{-6}$ eV, the ion convergence criterion is $5 \cdot 10^{-5}$ eV with respect to energy, and the projection is done in reciprocal space.
When the volume of the unit cell change, the plane wave basis set might not be as accurate as the starting settings.
To handle this, the unit cell workflow rerelaxes the structure until the energy difference between iterations is smaller than $5 \cdot 10^{-4}$ eV.

After the unit cell has been optimized, the supercell is created.
The supercell size is set to be approximately 20 Å in every direction as default to ensure a low defect-defect interaction. To preserve the symmetry of the crystal, the unit cell is copied until the size criterion is met.
To get the energy of the host supercell, it is processed in the host workflow that is similar to the ground state workflow (Sec.~\ref{sec:ground}) but only runs the first four steps but with symmetry turned on.
Then the point defect supercells, which will be denoted defect cells from now on, are generated as described in Appendix~\ref{appendix:defect_gen}.
Now, the defect cells can either be screened -- as described in Appendix~\ref{appendix:screen} -- or run through the full characterization workflow directly.

\section{Full characterization workflow}
\label{sec:workflow}

\begin{figure*}[h!]
   \includegraphics[width=\textwidth]{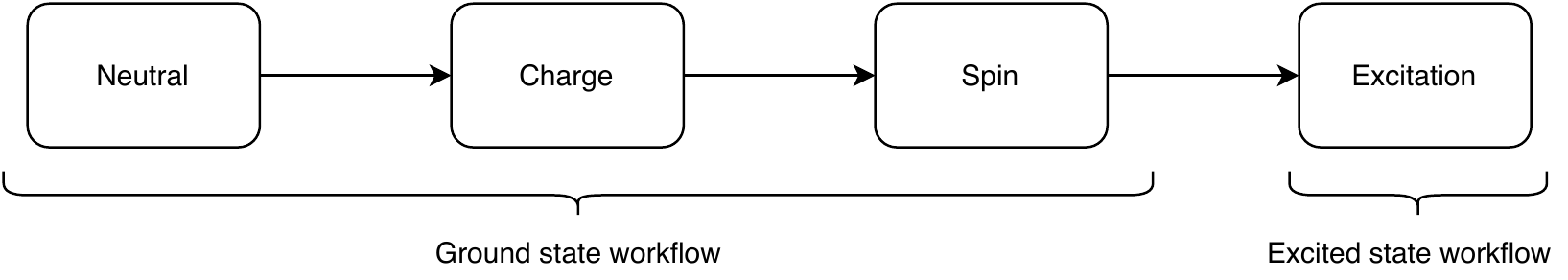}
	\caption{The general design of the full characterization workflow. It divides the calculations into four steps and carries them out in the order shown in the figure. The neutral, charge, and spin steps are processed with the ground state workflow (Sec.~\ref{sec:ground}), whereas the excitation steps are processed in the excited state workflow (Sec.~\ref{sec:excitation}).}
	\label{fig:full} 
\end{figure*}

After the host material has been selected, the unit cell relaxed, and the defect cells have been generated, the full characterization workflow can start.
Figure~\ref{fig:full} shows an overview of the different steps involved to fully characterize any point defect cluster.
Hereinafter, we refer to these steps as neutral, charge, and spin step, which include neutral and charged ground state calculations and alternative spin calculations, respectively.
These steps are processed with the ground state workflow, see Sec.~\ref{sec:ground}.
First, the neutral defect is processed.
After this step has finished and there are defect states present in the band gap, the charge step follows.
The charge step runs the ground state workflow again but now with charged supercells.
The default settings remove and add up to two electrons.
These charge runs, a maximum of four, are processed in parallel.
When the charge step is finished, alternative spin states are processed with the ground state workflow.
For this step, the ground state electronic occupation is changed (for example from spin-3/2 to spin-1/2), and the alternative spin states are calculated for the point defect.
The alternative spin step is carried out for both the neutral and the charged defects with a maximum of two alternative spins per charge state, usually, only one is found.
For five different charge states, usually, five different spin states are processed, also in parallel.
Additional details about the input for both charge and spin steps are presented in the ground state workflow in the analyze section, see Sec.~\ref{sec:analyze}.

Once the neutral, charge, and spin steps are finished, the excitation step starts.
This step calculates all the excitations, which is the most time-consuming step, and uses a separate workflow described in Sec.~\ref{sec:excitation}.
All possible single excitations are calculated for all stable ground state configurations obtained in neutral, charge, and spin steps.
The main focus is the local excitations between the defect states in the band gap.
However, as default, the excitations to and from the conduction and valence band are also included.
Due to the large number of excitations, a limit of the excitation energy is added as a default.
If the excitation energy is too small, it is difficult to observe due to the increased risk of non-radiative decay from the excited state to the ground state and because the experimental equipment often has a limit on how small photon energy can be detected.
We assume that the experimental limit for regular detectors is around 0.68 eV (1800 nm).
The threshold value is set to 0.4 eV, due to the systematic underestimation of the ZPL of 0.2 eV from the PBE functional as mentioned in the method section.
Any excitation below 0.4 eV is neglected.

\subsection{Ground state workflow}
\label{sec:ground}

\begin{figure*}[h!]
   \includegraphics[width=\textwidth]{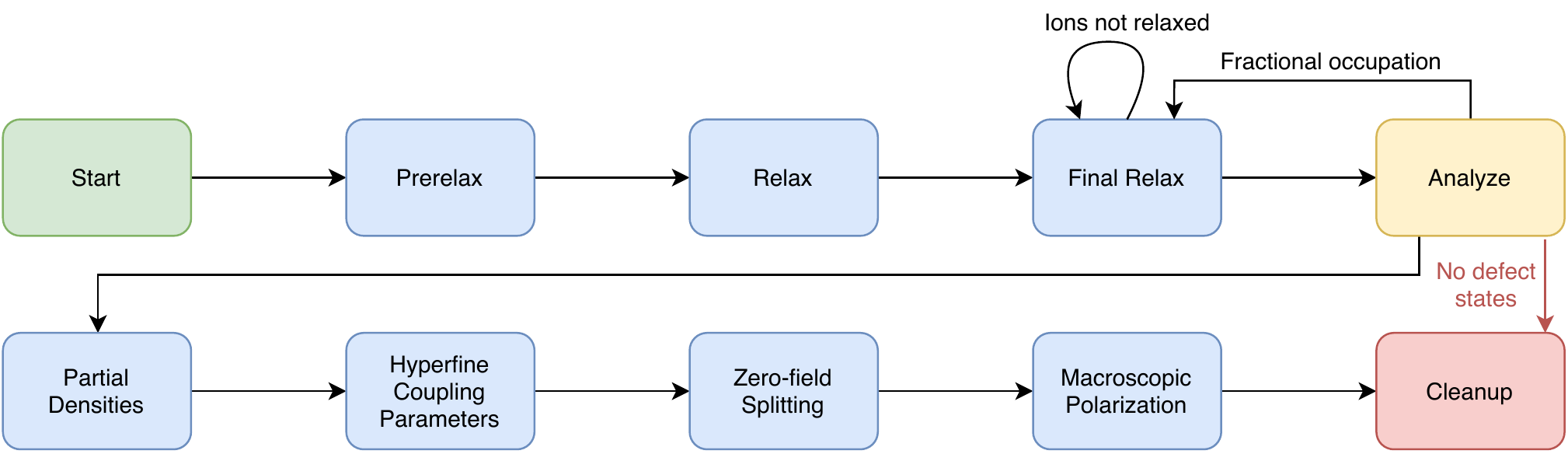}
	\caption{Flowchart of the ground state workflow that relaxes, analyzes, and post-processes a defect cell. Green and red boxes show start and end. The blue boxes show VASP calculations, and the yellow box shows the analyze step.}
	\label{fig:ground} 
\end{figure*}

Figure~\ref{fig:ground} shows an overview of the ground state workflow.
The first row shows the electronic and ionic relaxation calculations divided into three steps.
This division is done to optimize the convergence.
Compared with using one relaxation step with the highest settings, this division halves the computational time per relaxation.
The second row shows the post-processing steps.
The details for each step are presented in separate subsections below.

\subsubsection{Start}

First, the data files with structure and computational parameters are copied into the running directory, and \textit{httk} selects the PAW pseudopotentials.
For charge and spin calculations, the number of electrons and fixed occupation is added to the input, respectively.

\subsubsection{Prerelax}

This first step is a fast and coarse ion relaxation to ensure the ions are not too close.
Here, the convergence settings are reduced: electronic convergence criterion is $10^{-4}$ eV; ion convergence criterion is $0.3$ eV with respect to forces and includes 20 ionic relaxation steps; spin polarization is turned off; the FFT grid is set to 3/2 of the largest wave vector; and only the $\Gamma$-point is used for the integration over the Brillouin zone.
When the calculation is finished, the wave function is saved and used as an input for the next step.
Since this step only uses the $\Gamma$ point, it could be run with the gamma-only version of the VASP software.
However, the wave function from this version cannot be loaded into the standard version of VASP, and our tests show that the overall time is lower if the wave function is propagated from the prerelax to relax step.
This step is only carried out for the neutral defect.

\subsubsection{Relax}
\label{sec:relax}

Here, the charge density and wave function from the prerelax step is loaded.
The workflow starts here for charge and spin steps and loads the wave function from the neutral prerelax step.
In this step, the convergence parameters are increased: ion convergence criterion is $5 \cdot 10^{-4}$ eV with respect to energy and 30 ionic relaxation steps; spin polarization is turned on; the FFT grid is set to twice of the largest wave vector; and $2 \times 2 \times 2$ k-point grid is used.

\subsubsection{Final relax}
\label{sec:final}

In this final ion relaxation step, the electronic convergence criterion is increased to $10^{-6}$ eV, and the ion convergence criterion is changed $5 \cdot 10^{-5}$ eV, still with respect to forces and up to 30 ionic relaxation steps are included.
If the calculations fail to relax the ions during any of these three relaxation steps, the \textit{httk} checkers try to find and correct the error to converge the calculation in this final step.
If this is not possible, it aborts and proceeds to the cleanup step.
When the ion relaxation is completed, the final wave function is saved and analyzed in the next step.

\subsubsection{Analyze}
\label{sec:analyze}

First, if fractional occupation occurs, the nearest non-fractional occupation is forced, and the final relax step is repeated.
Next, one needs to identify if defect states are present in the band gap.
This identification uses the inverse participation ratio (IPR)~\cite{kramer1993localization,pashartis2017localization}, which is a measure of localization.
In practice, the IPR is calculated with a python library PyVaspwfc~\cite{zheng_2019} for the 30 bands closest to the Fermi level in each spin channel and averaged over all k-points.
If a band has an IPR larger than the threshold ($10^{-4}$), it is considered a defect state.
Sometimes, stray band appears in the valence or conduction band which goes above this threshold suggesting it may be a defect state.
The identified defect bands should be continuous, and thus any outliers are removed, these are saved in a separate file since they can provide further information about the point defect but cannot be handled in the workflow.
If no local states are found, the workflow ends.

Once the defect states have been identified, the input for the rest of the workflow can be constructed.
For the charge state calculations, the number and occupation of defect states are checked.
If there are empty defect states, additional electrons (maximum two) are added to the supercell.
Likewise, if there are filled defect states then electrons (maximum two) are removed.
The setup part for the spin state calculation is more complicated.
First, the algorithm double-checks that electrons are not taken from the valence band or moved into the conduction band by checking that the highest occupied state and the first unoccupied state are defect states in each spin channel.
If this is not the case in a spin channel, that channel is excluded.
Next, the algorithm checks if the defect is mirrored, \emph{i.e.}, exhibiting the same structure in both spin channels.
If the number of defect states and the highest occupied state is the same in both spin channels, then the defect is assumed to be mirrored, and one spin channel is excluded.
In this case, only one alternative spin exists, the output is set up, and the algorithm ends.
However, if both spin channels fulfill all these criteria, the final part of the algorithm moves electrons between both spin channels, producing two alternative spins.
In this case, the different spins are ordered in terms of rising energy by comparing how the different occupations would affect the eigenvalues in each spin channel.
The spin finding algorithm finishes, and the excitation algorithm takes over.

The final part in the analyze step sets up the excitations.
Three different kinds of excitations are handled: local (bound-to-bound), valence band (free-to-bound), and conduction band (bound-to-free).
The number of excitations depends on the number of local states (N) and the number of occupied states (O).
The number of local excitations ($\mathrm{N_{LE}}$) follows
\begin{equation}
\mathrm{N_{LE}}=\binom{N}{O}-1-d;
\end{equation}
where the first term is the binomial coefficient, the second term excludes the ground state, and the final term excludes any double excitation.
The number of valence band excitations ($\mathrm{N_{VE}}$) depends on the non-occupied states (N-O), and the number of conduction band excitations ($\mathrm{N_{CE}}$) depends on the occupied states (O), hence together, they scale as N.
The total number of excitations ($\mathrm{N_{TE}}$) would be $\mathrm{N_{TE}}=\mathrm{N_{LE}}+\mathrm{N_{VE}}+\mathrm{N_{CE}}=\binom{N}{O}-1-d+(N-O)+(O)=\binom{N}{O}-1-d+N$.
For example, a defect with 4 local states and 2 occupied states has $\mathrm{N_{TE}}=\binom{4}{2}-1-1+4=8$ excitations. 
These excitations are handled in a separate workflow, see Sec.~\ref{sec:excitation}.
After the analyze step is completed, the post-processing steps start.

\subsubsection{Partial densities}

The first post-processing step calculates and saves the partial densities for the local states as well as the top and bottom of the valence and conduction band, respectively.

\subsubsection{Hyperfine coupling parameters}
\label{sec:hyp}

If the spin of the defect and the nuclear spin of the chemical elements are non-zero, hyperfine tensors are calculated. 
The nuclear spin depends on the nuclear g-factor, which in turn depends on the isotope.
Only certain ions have a non-zero nuclear g-factor, for example, $^{13}$C has a non-zero value while $^{12}$C and $^{14}$C do not.
The g-factors are extracted from the easyspin website~\cite{easyspin}.
The algorithm calculates the hyperfine interaction for all possible paramagnetic isotopes.
An intrinsic defect in SiC has only $^{13}$C and $^{29}$Si with non-zero values, hence only one hyperfine interaction exists.
But if one would dope with B and N, which have non-zero values for both isotopes, then four hyperfine interactions would exist.
One calculation of the hyperfine coupling parameters is carried out in VASP with g-factors set to unity, and the results are multiplied with the different g-factor to get all the hyperfine interactions.

\subsubsection{Zero-field splitting}

If the spin of the defect is larger than one-half, the D-tensor is calculated as described in Ref.~\onlinecite{Ivady2014}.

\subsubsection{Macroscopic polarization}
\label{sec:macropollo}

The final post-processing step calculates the macroscopic polarization of the defect cell using Berry phase calculation.

\subsection{Excited state workflow}
\label{sec:excitation}

\begin{figure*}[h!]
   \includegraphics[width=\textwidth]{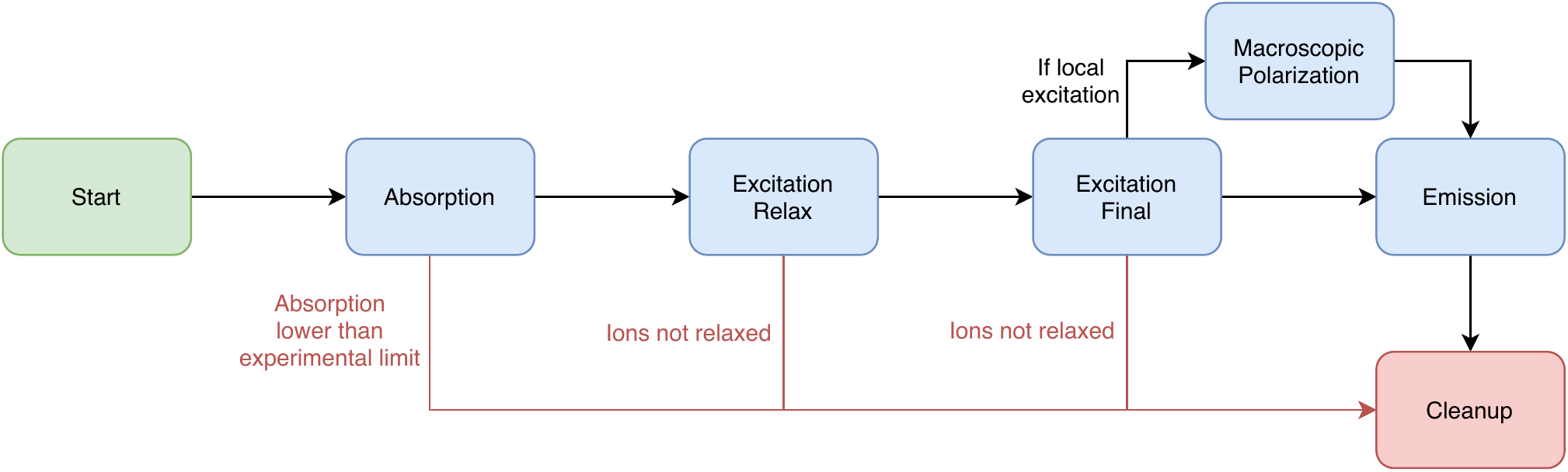}
	\caption{Flowchart of the excited state workflow that relaxes and post-processes a defect excitation. Green and red boxes show start and end. The blue boxes show VASP calculations.}
	\label{fig:excitation} 
\end{figure*}

After the ground state workflow is finished and the different excitations are set up, see Sec.~\ref{sec:analyze}, all excitations (local, valence, and conduction) for the different charge and spin states are processed in parallel.
Figure~\ref{fig:excitation} shows an overview of the separate workflow for the excited states.

The excitation calculations are more intricate to run than the ground state because the fixed electronic excitations may not have stable ion positions.
In this case, the calculation never converges.
To handle this, an additional checker stops any run if any relaxation step takes too long to relax.
This checker monitors the electronic iterations during each ion relaxation and saves the lowest number of electronic iterations needed to converge one electronic cycle.
If the number of electronic iterations goes above this number with a buffer of five steps, the run is stopped.

\subsubsection{Absorption}

The first step of this workflow calculates the absorption energy  by setting the excited state occupation and running an electronic cycle without relaxing the ionic positions.
This step starts from the final relaxed ground state (Sec.~\ref{sec:final}) wave function and does 60 electronic iterations or until the electronic convergence criterion of $10^{-6}$ is reached.
If it fails to converge during these 60 iterations or if the final energy difference between this state and final relax ground state is lower than the experimental limit (0.4 eV, see Sec.~\ref{sec:method}), the run is stopped, and the workflow proceeds to the cleanup step.

\subsubsection{Excitation relax}

In this step, the same settings are used as in the relax step in the ground state workflow, see Sec.~\ref{sec:relax}, but with excited state occupation.

\subsubsection{Excitation final relax}
\label{sec:ex_final}

In this step, the same settings are used as in the final relax step in the ground state workflow, see Sec.~\ref{sec:final}, but with excited state occupation.

\subsubsection{Macroscopic polarization}

If the excitation is local, the macroscopic polarization of the excited state is calculated the same way as in Sec.~\ref{sec:macropollo}.

\subsubsection{Emission}

When the final excited state geometry is found, the occupation is set to the ground state occupation.
This step is the counterpart to the absorption step; here, the excited state geometry is used with the ground state occupation.

\subsubsection{Post-processing}
\label{sec:post_pro}

Between each step in the excited state workflow, the following post-processing steps take place.
The partial charge densities are calculated the same way as in the ground state workflow.
Here, the TDM is also calculated in two ways.
First, between each step in the excited state workflow, for example, from absorption to excitation relax step.
Second, between the excited state step and its corresponding ground state step, for example, from excitation relax step in excited state workflow to relax step in ground state workflow.
To calculate the TDM, the WAVECARs are post-processed using the PyVaspwfc python library~\cite{zheng_2019}, which we have modified to handle two WAVECARs and calculate the TDM between the defect orbitals.

\subsection{Cleanup}
\label{sec:clean}

During the full characterization workflow run, several WAVECARs are saved for optimization or post-processing purposes.
For our 4H-SiC example, each wavecar is 40 Gb, for five charge states with alternative spin states, the total is 400 Gb for the ground states.
For an excited state run, the WAVECARs are also saved during the run.
Since the post-processing step takes the difference between the current and previous steps, on average, there are two WAVECARs saved per excitation.
If ten taskmanagers are running in parallel, the total storage quickly goes up to 1 Tb.
After all steps in both the ground and excited state workflows are completed, the WAVECARs are deleted, and the remaining files are compressed.
The final storage after removing the WAVECARs is about 50 Gb.
When the full characterization workflow is finished, about 500 VASP calculations have been performed.

\section{Database}
\label{sec:database}

After the full characterization workflow has finished, all the properties described in the theory section (Sec.~\ref{sec:theory}) are collected and stored in a database.
Additional outputs include the bands closest to the Fermi energy with the corresponding localization value derived from IPR.

\subsection{Photoluminescence}

The energy at every step in the excitation cycle is saved to the database.
The ZPL is the total energy difference between the final relax ground state calculation (Sec.~\ref{sec:final}) and final relax excited state calculation (Sec.~\ref{sec:ex_final}).
Similarly, the absorption and emission are calculated.
For these two properties, the only difference is that the ion positions are the same in both the ground and excited states.
Note that the TDM is saved between all defect states throughout the excitation cycle.
From the TDM, the polarization and lifetime are estimated and saved to the database.
The $\Delta Q$, Eq.~\eqref{eq:dq}, is also calculated and saved.

\subsection{Hyperfine coupling parameters and ZFS}

For the hyperfine interaction, the three different tables related to hyperfine coupling parameters, as described in the VASP manual~\cite{vaspwiki}, are extracted from the run and post-processed.
Since the calculations are run with the g-factors set to unity, each atom is multiplied with the corresponding g-factor to produce all hyperfine interactions (Sec.~\ref{sec:hyp}).
For each atom, the bipolar hyperfine coupling parameter matrix is extracted, and the Fermi contact coupling parameter ($\mathrm{A}_{1\mathrm{c}}$ and $\mathrm{A}_\mathrm{tot}$) diagonal matrix is constructed.
The eigenvalues and eigenvectors of these added matrices are calculated as well as the eigenvectors' angles to the z-axis.
Since symmetrical equivalent atoms have the same hyperfine coupling parameters, the multiplicity of each hyperfine value is also marked.
The hyperfine coupling parameters are only saved if the hyperfine splitting ($\mathrm{A}_\mathrm{z}$) is larger than 3 MHz.

The D-tensor is extracted from the output, both the calculated and diagonalized version with corresponding eigenvectors.

\subsection{Bands with IPR}

The 30 bands closest to the highest occupied band in both spin channels are saved to the database.
This includes the local bands, if any exist, and a few bands in the VB and CB, respectively.
For easy identification of local bands, the IPR value (discussed in detail in Sec.~\ref{sec:analyze}) is also stored for each band and both spin channels.

\subsection{Formation energy and charge-state transition levels}

Formation energy for all charge and spin states as well as charge-state transition levels between all different charge and spin states are calculated and stored.

\section{Computational results}
\label{sec:resuls}

To demonstrate the results produced by ADAQ, the silicon vacancy ($\mathrm{V}_{\mathrm{Si}}$) in 4H-SiC is used as an example.
SiC is a technologically mature material where it is possible to combine quantum and classical applications in the same device~\cite{castelletto2020silicon}.
However, it is a material with many different polytypes and multiple unidentified defects.
The silicon vacancy has two configurations in 4H-SiC, denoted $h$ and $k$.
Given that these configurations have been identified earlier~\cite{castelletto2020silicon}, both these configurations are processed directly through the full characterization workflow.
If these had been unknown defects, the best approach would be to generate an array of different point defect clusters and run these defects through the screening workflow (Appendix~\ref{appendix:screen}), a scaled-down version of the full characterization workflow.
After this step, the most likely candidates are processed through the full characterization workflow to identify the unknown defects.
We present data for the silicon vacancy from both the full characterization and screening workflow, starting with the former.

\clearpage

\subsection{Full characterization workflow results}

\begin{figure}[h!]
   \includegraphics[width=\textwidth]{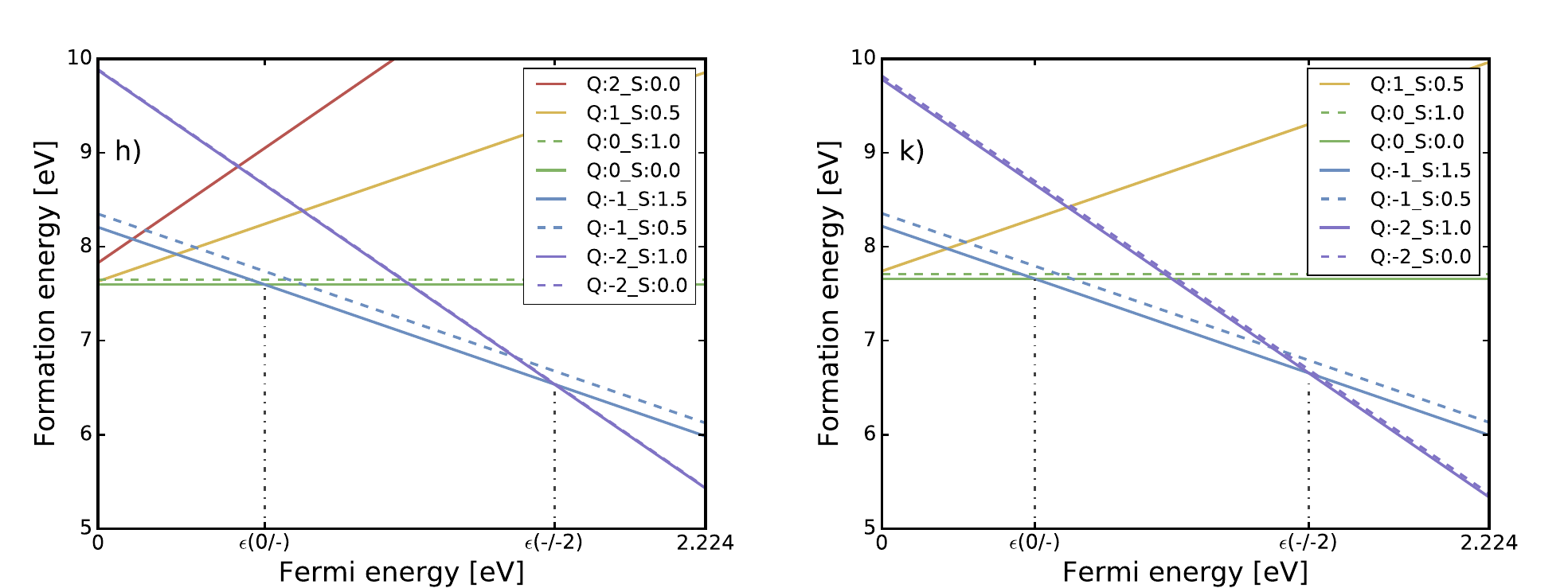}
	\caption{The formation energy and charge transition levels for the $h$ and $k$ configurations of the $\mathrm{V}_{\mathrm{Si}}$ in 4H-SiC. The spin state configurations with the lowest energy are plotted with solid lines, the others with dashed. The labels denote the charge (Q) and spin (S) states of the defect. The x-axis is the Fermi energy which starts at zero and goes up to the band gap energy. Both configurations have two charge-state transition levels, one from neutral to negative and one from negative to double negative.}
	\label{fig:formation_energy} 
\end{figure}

First, Figure~\ref{fig:formation_energy} shows the formation energy with charge transition levels for both configurations of the $\mathrm{V}_{\mathrm{Si}}$, with the chemical potential for the silicon: $\mu_{\mathrm{Si}}^{\mathrm{rich}}=-5.42$ eV.
Using the charge correction formula in Eq.~\eqref{eq:cc} with the values $(1+f)=0.65$, $\alpha_M=2.8373$ (Madelung constant for simple cubic~\cite{leslie1985energy} since the 576 supercell is close to cubic), $\epsilon_r=9.6$ (taken from experiment~\cite{willardson1998sic}), and $L=V^{1/3}=18.21\ \mathrm{\AA}$ for the 576 atom supercell gives a correction of 0.076 eV for $q=\pm1$ and 0.304 eV for $q=\pm2$.

\begin{figure}[h!]
   \includegraphics[width=\textwidth]{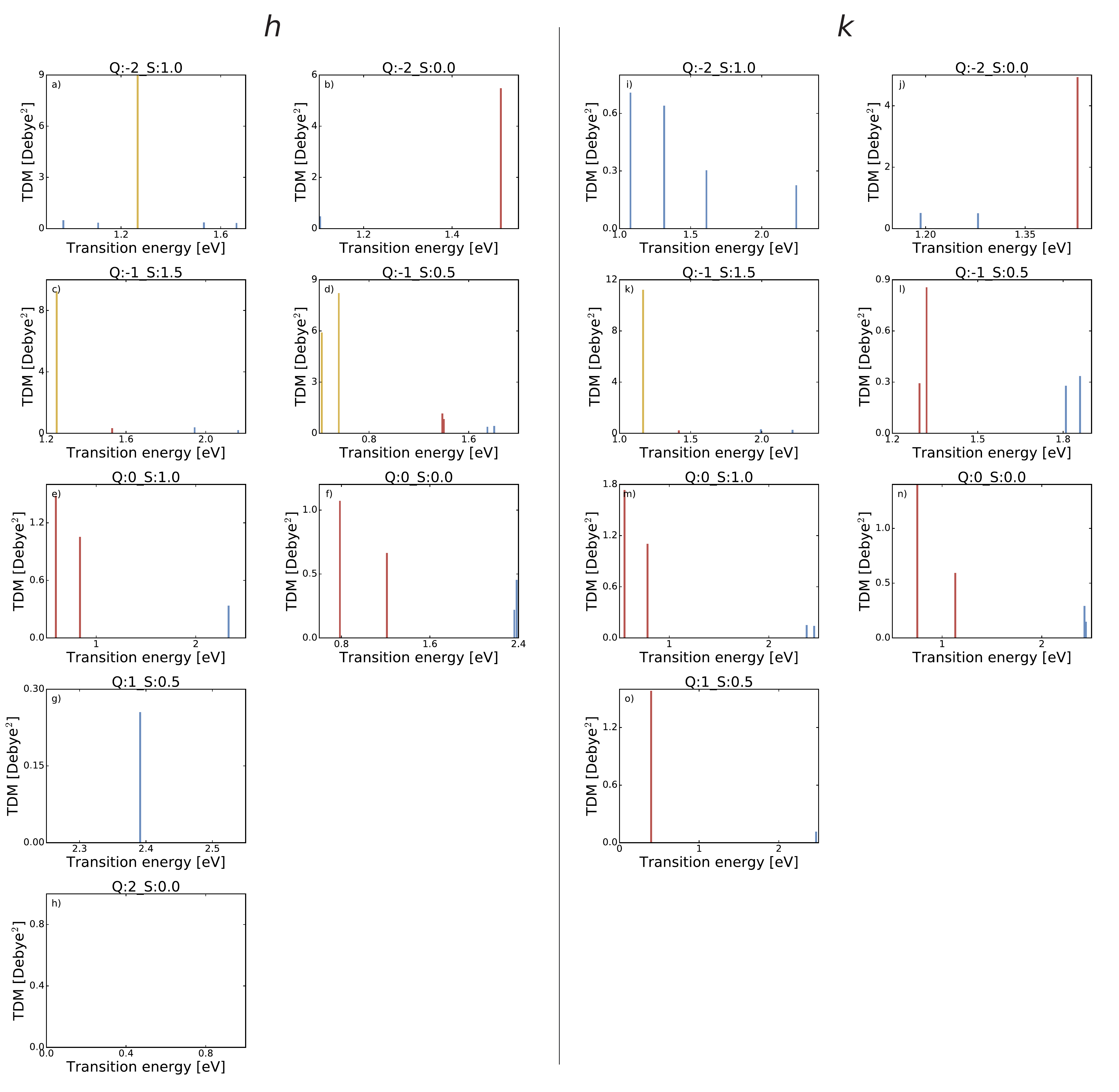}
	\caption{Spectral lines for all charge and spin states found by the full characterization in ADAQ for the $\mathrm{V}_{\mathrm{Si}}$ in 4H-SiC. For the $h$ configuration, a) and b) are the double negative states, c) and d) are the negative states, e) and f) are the neutral states, g) is the positive state, and h) the double positive state. For the $k$ configuration, i) and j) are the double negative states, k) and l) are the negative states, m) and n) are the neutral states, and o) is the positive state. The yellow lines are ZPL (bound-to-bound transition); the blue lines are defect state to conduction band edge (bound-to-free transition); and the red lines are valence band edge to defect state (free-to-bound transition). The x-axis is the transition energy and the y-axis is the TDM intensity $|\bar{\boldsymbol{\mu}}|^2$.}
	\label{fig:state} 
\end{figure}

Next, the spectral lines for all the charge and spin states for both configurations of the $\mathrm{V}_{\mathrm{Si}}$ in 4H-SiC are presented in Figure~\ref{fig:state}.

\clearpage

\clearpage

\begin{figure}[h!]
   \includegraphics[width=\textwidth]{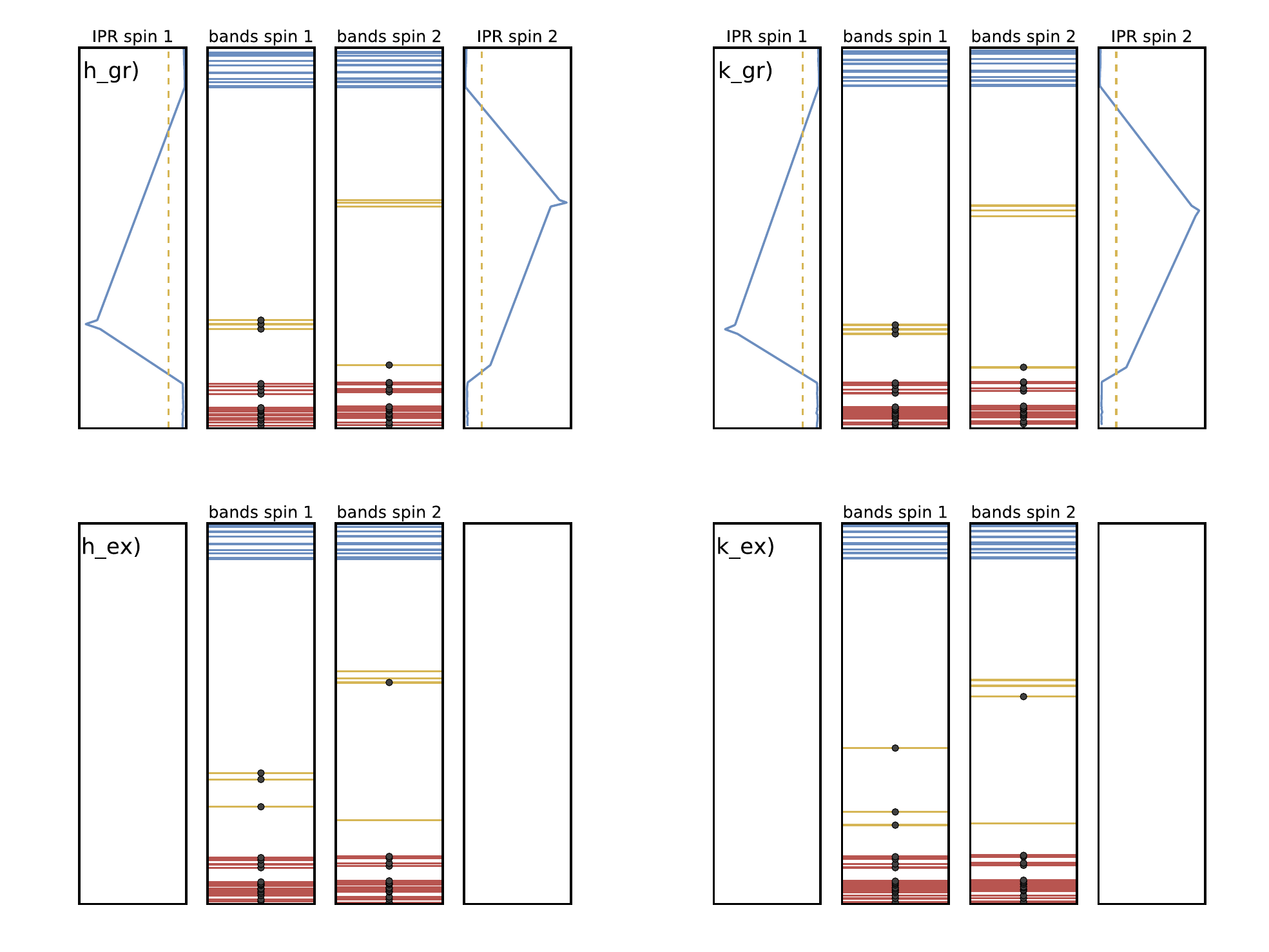}
	\caption{Kohn-Sham electronic structure for the negatively charged spin-3/2 configurations of the $\mathrm{V}_{\mathrm{Si}}$ in 4H-SiC. h\_gr) and k\_gr) show the ground state with IPR whereas h\_ex) and k\_ex) show the excited state. In the middle are the Kohn-Sham eigenvalues for the spin 1 and spin 2 channels. Blue and red bands denote conduction and valence bands, respectively. Yellow bands denote defect bands. On each side is the IPR with the threshold used in the full characterization workflow to identify the local bands. The black dots show the occupation.}
	\label{fig:band_data} 
\end{figure}

Looking at Figure~\ref{fig:state}, only the negative charge state with spin-3/2 has a ZPL for both configurations of the $\mathrm{V}_{\mathrm{Si}}$.
For this charge-spin state, the Kohn-Sham electronic structure is presented in Figure~\ref{fig:band_data} for both configurations.
Here, the 30 closest bands to the Fermi level are presented for both ground and excited states, as well as the IPR values for the ground state.
Since the full characterization workflow does not calculate IPR for the excited state, these are left blank.
Both the eigenvalues and IPR are averaged over the k-points.
The defect bands in the band gap are identified as local bands.
In the spin 2 channel, one local state close to the valence band just passes the localization limit.
In the excited state, this band moves up into the band gap.

\clearpage

For the negatively charged spin-3/2 states, Table~\ref{tab:optical} shows the polarization and lifetime of the ZPL.
The polarization changed between the different steps in the excitation cycle, most visible between the absorption and ZPL.
The ion relaxation factors for the $h$ configuration are as follows: $\Delta R= 0.16\ \mathrm{\AA} $ and $\Delta Q = 0.60\ \mathrm{amu}^{1/2}\mathrm{\AA}$. 
For the $k$ configuration $\Delta R = 0.18\ \mathrm{\AA} $ and $\Delta Q = 0.74\ \mathrm{amu}^{1/2}\mathrm{\AA}$.

\begin{table}[h!]
\caption{Polarization, intensity, and lifetime for the negatively charged spin-3/2 states of the $\mathrm{V}_{\mathrm{Si}}$ in 4H-SiC.}
\begin{ruledtabular}
\begin{tabular} {l|l|rrrr}
Configuration & Data & Absorption & ZPL & Emission \\ \hline
\multirow{6}{*}{$h$} & Energy [eV] & 1.324 & 1.252 & 1.173\\
& $\bar{\mu}_x$ [Debye] & 4.451 & 5.833 & 5.493 \\
& $\bar{\mu}_y$ [Debye] & 5.187 & 4.616 & 3.382\\
& $\bar{\mu}_z$ [Debye] & 3.913 & 5.386 & 4.784\\
& $|\bar{\boldsymbol{\mu}}|^2$ [Debye$^2$] & 62.03 & 84.34 & 64.50\\
& $\tau$ [ns] & 15.94 & 13.87 & 22.02 \\
\hline
\multirow{6}{*}{$k$} & Energy [eV] & 1.275 & 1.166 & 1.033\\ 
& $\bar{\mu}_x$ [Debye] & 3.978 & 0.713 & 0.464\\
& $\bar{\mu}_y$ [Debye] & 4.601 & 1.732 & 1.201\\
& $\bar{\mu}_z$ [Debye] & 5.402 & 10.98 & 10.19\\
& $|\bar{\boldsymbol{\mu}}|^2$ [Debye$^2$] & 66.17 & 124.0 & 105.5\\
& $\tau$ [ns] & 16.74 & 11.68 & 19.74\\ 
\end{tabular}
\end{ruledtabular}
\label{tab:optical}
\end{table}

Additional data about hyperfine coupling parameters and ZFS for the negatively charged spin-3/2 states are presented in Table~\ref{tab:hyp} and Table~\ref{tab:zfs}.
For both configurations of the $\mathrm{V}_{\mathrm{Si}}$, the first row is the largest hyperfine splitting related to the carbon atom above the silicon vacancy, and the second row is the three carbons below the silicon vacancy.
Both the hyperfine coupling parameters and ZFS show a small but perceivable difference between the two configurations.
These results and trends are comparable with HSE06 hyperfine and ZFS data~\cite{ivady2017identification}.

\begin{table}[h!]
\caption{Hyperfine tensor for the negatively charged spin-3/2 configurations state of the $\mathrm{V}_{\mathrm{Si}}$ in 4H-SiC. $\mathrm{A_{xx}}$, $\mathrm{A_{yy}}$, $\mathrm{A_{zz}}$, and $\mathrm{A_{z}}$ are given in MHz and the angles in degrees.}
\begin{ruledtabular}
\begin{tabular} {l|l|rrrr}
Configuration & Nucleus(multiplicity) & $\mathrm{A_{xx}(\theta)}$ & $\mathrm{A_{yy}(\theta)}$ & $\mathrm{A_{zz}(\theta)}$ &  $\mathrm{A_z}$ \\ \hline
\multirow{7}{*}{$h$} & C(1) & 22.35 (89.99) & 22.35 (90.0) & 71.25 (0.01) & 71.25 \\
& C(3) & 20.3 (18.91) & 20.41 (90.0) & 70.11 (71.09) & 29.75 \\
& C(3) & 3.4 (89.68) & 3.49 (72.36) & 7.04 (17.65) & 6.79 \\
& C(3) & 3.14 (90.01) & 3.22 (36.39) & 6.95 (53.61) & 4.87 \\
& C(6) & 3.57 (40.34) & 3.64 (57.27) & 7.46 (75.96) & 4.01 \\
& Si(3) & 5.72 (74.82) & 6.64 (15.37) & 6.67 (88.07) & 6.58 \\
& Si(9) & 5.32 (50.5) & 5.97 (49.71) & 6.02 (76.93) & 5.74 \\
\hline
\multirow{8}{*}{$k$} & C(1) & 23.08 (90.02) & 23.08 (90.0) & 70.23 (0.02) & 70.23 \\
& C(3) & 17.24 (19.74) & 17.42 (89.39) & 66.92 (70.28) & 27.81 \\
& C(3) & 3.93 (89.78) & 4.05 (69.11) & 7.77 (20.9) & 7.4 \\
& C(6) & 2.75 (10.44) & 2.88 (80.9) & 6.16 (85.06) & 4.41 \\
& Si(6) & 5.97 (57.88) & 6.49 (33.3) & 6.52 (82.28) & 6.35 \\
& Si(3) & 5.42 (41.56) & 6.57 (89.87) & 6.7 (48.44) & 6.02 \\
& Si(3) & 5.19 (80.4) & 5.77 (87.99) & 5.82 (9.91) & 5.81 \\
& Si(1) & -4.82 (0.0) & -4.33 (90.0) & -4.33 (90.0) & 4.82 \\
\end{tabular}
\end{ruledtabular}
\label{tab:hyp}
\end{table}

\begin{table}[h!]
\caption{Zero field splitting tensor for the negatively charged spin-3/2 configurations of the $\mathrm{V}_{\mathrm{Si}}$ in 4H-SiC. The tensor and diagonal are given in MHz and the eigenvalues are unit vectors.}
\begin{ruledtabular}
\begin{tabular} {l|c|cc}
Configuration & Tensor & Diagonal & Eigenvalues (x,y,z) \\ \hline
$h$ & $\begin{pmatrix}
-8.716 & 0.004 & -0.000\\
0.004 & -8.509 & -0.135\\
-0.000 & -0.135 & 17.225\\
\end{pmatrix}$  & $\begin{pmatrix}
-8.510 & 0 & 0\\
0 & -8.716 & 0\\
0 & 0 & 17.226\\
\end{pmatrix}$ & $\begin{pmatrix}
0.020  &   1.000  &   0.005\\
1.000  &  -0.020  &  -0.000\\
-0.000 &   -0.005 &    1.000\\
\end{pmatrix}$  \\
$k$ & $\begin{pmatrix}
-13.479 & -0.100 & 0.024\\
-0.100 & -13.577 & -0.010\\
0.024 & -0.010 & 27.057\\
\end{pmatrix}$ & $\begin{pmatrix}
-13.417 & 0 & 0\\
0 & -13.640 & 0\\
0 & 0 & 27.057\\
\end{pmatrix}$ & $\begin{pmatrix}
-0.849 & 0.529 & 0.001\\
0.529 & 0.849 & -0.000\\
0.001 & -0.000 & 1.000\\
\end{pmatrix}$ \\ 
\end{tabular}
\end{ruledtabular}
\label{tab:zfs}
\end{table}

\clearpage

\subsection{Screening workflow results}

As a demonstration of the screening workflow described in Appendix~\ref{appendix:screen}, the silicon vacancy is also processed through this workflow and the results are presented in Table~\ref{tab:screen}.
Eigendifference stands for eigenvalue difference.

\begin{table}[h!]
\caption{Ground state energy and point defect spin as well as ZPL and TDM estimates from the screening workflow for the $\mathrm{V}_{\mathrm{Si}}$ in 4H-SiC. If two charge states with different spin are presented, the one with the lowest ground state energy is presented first.}
\begin{ruledtabular}
\begin{tabular} {l|rr|rrrr}
 & \multicolumn{2}{c|}{Ground state properties}  & \multicolumn{4}{c}{Excited state properties} \\
Configuration &  &  & \multicolumn{2}{c}{Smallest transition} & \multicolumn{2}{c}{Transition of interest} \\ \cline{4-5} \cline{6-7}
(charge state) & Energy [eV] & Spin & Eigendifference [eV] & TDM [Debye] & ZPL [eV] & TDM [Debye] \\ \hline
h(-1) & -4315.19047813 & 3/2 & 1.41 & 8.70 & 1.27 & 10.48 \\
h(-1) & -4315.04218904 & 1/2 & 0.66 & 7.90 & 0.60 & 5.35 \\
h(0) & -4323.75749323 & 0 & 0.65 & 4.97 & 0.29 & 4.41 \\
h(0) & -4323.71386418 & 1 & 0.52 & 2.51 & 0.15 & 4.85 \\
h(+1) & -4331.91195768 & 1/2 & 0.24 & 5.35 & 0.14 & 4.68 \\
\hline 
k(-1) & -4315.18187477 & 3/2 & - & - & - & -  \\
k(-1) & -4315.03110113 & 1/2 & 0.67 & 3.41 & 0.37 & 4.84 \\
k(0) & -4323.69812683 & 0 & 0.64 & 3.97 & 0.27 & 4.40 \\
k(0) & -4323.65815249 & 1 & 0.52 & 1.35 & 0.13 & 1.72 \\
k(+1) & -4331.79353495 & 1/2 & 0.24 & 5.54 & 0.14 & 4.28 \\
\end{tabular}
\end{ruledtabular}
\label{tab:screen}
\end{table}

\section{Discussion}
\label{sec:discussion}

In this section, we discuss the strengths and limitations of the screening and full characterization workflows, starting with the latter.

\subsection{Full characterization workflow comments}

We start with the formation energy in Figure~\ref{fig:formation_energy}.
Comparing this figure with Figure 3 in Ref.~\onlinecite{Szasz2015}, which was calculated with HSE functional, the most obvious difference is the band gap.
Indeed, the use of the PBE functional in the ADAQ workflows gives a band gap of 2.224 eV for 4H-SiC, whereas the HSE calculations in Ref.~\onlinecite{Szasz2015}, gives a band gap about 3.2 eV, which is close to the experimental value of 3.23 eV~\cite{levinshtein2001properties}.
This difference shifts the charge transition levels, but overall, the formation energy curves have the same shape.
Another effect on the charge transition levels is the choice of the charge correction scheme.
In this article, the LZ correction is used, whereas the FNV correction was used in Ref.~\onlinecite{Szasz2015}.
However, the differences are minor.
An additional triple negative charge state was considered in Ref.~\onlinecite{Szasz2015}, which is not included in the workflow.
Both the use of the PBE functional and the restriction to five charge states are reasonable limitations needed in terms of efficiency for the full characterization workflow. 
Therefore, the charge transition level diagram obtained from ADAQ gives reliable information on the stability of defect charge states and allows one to determine the most stable charge and spin state.

When it comes to finding ZPLs for a point defect cluster, the full characterization workflow in ADAQ is an ideal choice.
It removes a lot of the guesswork of trying to estimate which charge and spin state of a point defect could produce a ZPL.
For example, the experimental lines 1.352 and 1.438 eV in 4H-SiC are related to the silicon vacancy with the negative charge state, and with the full characterization workflow, one gets both the charge and spin state of the defect.
The negative charge spin-3/2 state is the only state with a ZPL for both configurations in the reported experimental range, $k$ configuration is 1.166 eV and $h$ configuration is 1.240 eV.
Note that these ZPL calculated with the PBE functional are systematically underestimated by 0.2 eV compared with experiment but leaves the order unaffected~\cite{methodology_paper}.
Adding this 0.2 eV correction to the calculated results brings the ZPL almost on top of the experimental values, and the given order ($k$ is lower than $h$) agrees with previous identification~\cite{ivady2017identification}.
When comparing the corrected ZPL results from ADAQ with experiment, one gets an accurate picture of which ZPL belongs to which charge and spin state of the defect.
This is what ADAQ is designed to do.

Other useful data, which can help experimentalists, are the other lines presented in Figure~\ref{fig:state}.
The tiny red line close to 1.4 eV in Figure~\ref{fig:state}~k) shows valence band edge to defect state transition, corresponding to the defect affinity energy. 
It gives the limit of the laser energy used to excite the electron.
If an excitation energy larger than 1.6 eV (1.4+0.2 eV) is used, the defect will change to a double negative charge state as discussed in Sec.~\ref{sec:iae}, thus removing the visible ZPL from the spectra.
The 0.2 eV correction may be off for the free-to-bound and bound-to-free transitions since is estimated for bound-to-bound transitions.

Why can we neglect the charge correction for the ZPL calculations?
For the ZPL, there is a small multipole change between the ground and excited state. 
This multipole change has a minuscule effect on the ZPL energy.
For the $h$ configuration using the FNV correction in both the ground and excited state cause only a 5 meV difference in the ZPL.
This energy change is calculated between two local states, whereas the maximum multipole change comes from a local to delocalized transition.
An excitation from a local state to the conduction band minimum gives a slightly larger correction of 25 meV.
These corrections are still smaller than other uncertainties; hence the charge correction for the ZPL can be neglected.

An analysis of the band gap states for the negatively charged spin-3/2 state plotted in Figure~\ref{fig:band_data} can provide additional information.
In this plot, the defect band just above the VBM is sufficiently localized to be identified as a defect state.
The identification of this band as a local state occurs for this charge and spin state for both the $h$ and $k$ configurations.
For the $h$ configuration, this band is also a local state for the double negatively charged spin-1 state.
For all other charge and spin states, this band moves into the VB and mixes with the delocalized states.
In the excited state, this band moves up in the band gap, further supporting its identification as a local state.
The change is most significant at the $\Gamma$-point.
The IPR is not calculated for the excited state, mainly because it is not needed in the workflow. 
However, manual testing on the divacancy defect in 4H-SiC does not show any discernible difference in localization between the ground and excited state: the excited state IPR is almost identical compared to the ground state IPR.
The ZPL of interest is the transition between this band and the nearest higher energy band.
If the band closest to the VBM had not been identified as a defect state, the ZPL would have been a free-to-bound transition.
One should be aware that this might happen but as a precaution, the full characterization workflow already calculates the excitations to and from the valence and conduction band edges.

The transition energies with polarization, intensity, and lifetime are displayed in Table~\ref{tab:optical}.
Here, one can see all the energies in the excitation cycle, keeping in mind the systematic shift of 0.2 eV for ZPL.
The absorption and emission should have a similar shift.
The calculated lifetimes overestimate the experimental values somewhat.
For the $h$ configuration (V1) the lifetime is $5.5\pm1.4$ ns at 4.1 K~\cite{nagy2018quantum}.
One reason why the value (13.87 ns in Table~\ref{tab:optical}) for this configuration is too large could be the use of the PBE ZPL energy to calculate it.
This ZPL is underestimated by 0.2 eV.
Using the experimental ZPL (1.438 eV~\cite{sorman2000silicon}) would reduce the lifetime to 9.15 ns.
However, this is still larger than the experimental result.
The TDM may be slightly affected by the use of the PBE functional, using the HSE functional may increase the TDM further.
Looking at the $k$ configuration, $|\bar{\boldsymbol{\mu}}|^2$ is almost twice as large as for the $h$ configuration.
On the other hand, from Figure~1b) in Ref.~\onlinecite{nagy2018quantum}, the opposite should be the case.
These configurations are Jahn-Teller unstable, which also can affect the intensities.
Since the full characterization workflow uses k-points, the Jahn-Teller effects may not be correctly described, which could affect the polarization.
The $h$ configuration should have a ratio of 1.85 between the parallel (to the c-axis) and perpendicular components of the TDM~\cite{nagy2018quantum}.
But in this work, the value is 0.52 for $h$ configuration and 2.45 for the $k$ configuration.
Further accurate calculations are needed before any conclusion can be made about the polarization and intensity.
However, the lifetimes are reasonable, which can be used to identify promising defects for further in-depth analysis and more accurate, but time-consuming calculations.

The ion relaxation factors $\Delta R$ and $\Delta Q$ are similar for both configurations.
Since these values are small for the silicon vacancy, it suggests that the defect may be a good single photon emitter because of the spectral stability.
These values are also close to the values for the NV-center in diamond~\cite{alkauskas2014first}.

The fact that the negatively charged spin-3/2 state is responsible for the ZPL observed in experiment can further be supported by the hyperfine coupling parameters and ZFS.
For the hyperfine splitting, the $h$ configuration has slightly larger splitting for both the carbon above and the three carbons below the silicon vacancy than the $k$ configuration.
This agrees with the experimental findings in Ref.~\onlinecite{ivady2017identification}.
The theoretical results are about 10 Mhz lower than the experimental values.
The underestimation is due to the choice of the PBE functional, which typically overestimates the delocalization of the orbitals.
It has been demonstrated that the HSE functional gives hyperfine splitting results about 5 Mhz higher than the experimental values~\cite{ivady2017identification}.
The largest hyperfine splitting for silicon atoms should have a multiplicity of 12.
However, since the symmetry is off and dynamic Jahn-Teller is neglected, these multiplicities split into two different degeneracies of groupings of 9 and 3 for the $h$ configuration and three different degeneracies of groupings of 6,3, and 3 for the $k$ configuration.
Considering the ZFS, the $h$ configuration has a lower ZFS than the $k$ configuration, which also agrees with experiment~\cite{ivady2017identification}.
Overall, the data produced by the full characterization workflow gives a clear picture of the negatively charged spin-3/2 state of the silicon vacancy.

\subsection{Screening workflow comments}

Let us next discuss the accuracy of the results from the screening workflow, described in Appendix~\ref{appendix:screen}, in comparison to the full characterization workflow.
Overall, the ground state properties presented in Table~\ref{tab:screen} agree with the results for the full characterization workflow, shown in Figure~\ref{fig:formation_energy}.
The different spin states for the same charge state have the same order.
For the lowest negatively charged spin-3/2 state, the ground state energy of the $h$ configuration is 8.6 meV lower than for the $k$ configuration in the screening workflow. 
A similar difference is found in the full characterization (11.2 meV).
Even though the electronic and ionic convergence settings are reduced in the screening workflow, the ground state energies are accurate enough to determine which configuration and spin state is lowest in energy.

For the excited state properties, the ZPL and TDM results are produced faster but at a cost of accuracy.
First, there is a missing ZPL prediction for the spin-3/2 $k$ configuration due to the defect band closest to the VB edge did not have a high enough IPR to be classified as a localized state.
This misidentification is either because of the lower convergence settings or the smaller k-points grid, since this is not the case in the full characterization workflow as has been discussed in detail in a previous paragraph.
Second, the eigendifference seems to vary too much to make any accurate conclusions.
For the negatively charged $h$ spin 3/2 configuration, the eigendifference (1.41 eV) is surprisingly close to the experimental value (1.438 eV), the same can be observed for the divacancy in Figure~\ref{fig:comparison}, suggesting that the ion relaxation effect and the 0.2 eV underestimation cancel each other out.
But for the neutral $h$ configuration, this is not the case as comparing the eigendifference (0.65 eV) to the ZPL (0.29 eV), the difference is larger than 0.2 eV.
Hence, one must include the ion relaxation effects as the eigendifference can differ substantially even for the same defect in different charge states.
Third, the ZPL result from the screening workflow (1.27 eV) is close to the ZPL result from the full characterization workflow (1.25 eV).
Unfortunately, this is not always the case, as seen for the divacancy in Figure~\ref{fig:comparison} where the difference is about 0.1 eV.
Hence, the ZPL results from the screening workflow are within a $\pm$0.1 eV range compared to the full characterization workflow results.
 
\subsection{General workflow design comments} 
 
Finally, we provide some comments on the design and functionality of the workflows.
Overall, the automatic workflows present in ADAQ are powerful tools to find new interesting point defects.
They are consistent and do not require any human interference during the running of the workflows; hence, the risk of human error is minimized.
The number of calculations ($\sim$500 per defect) processed in the full characterization work also shows the need for automation.
With minimal manual overhead, the workflows produce data sufficient to identify the defect configurations and often provide relevant quantitative values for experimentally relevant parameters.
However, there is always a risk that one misses something when running high-throughput calculations.
In the example considered in this work, the screening workflow missed a transition for one of the silicon vacancy configurations.
Unfortunately, this is a side effect of the lower convergence settings necessary to screen a vast number of defects.
This risk of missing a transition can be minimized by screening all configurations of a defect and checking the results for consistency between the configurations.
On the other hand, the ground state energy from the screening workflow is accurate enough.
Therefore one can find the most stable defect which should be processed through the full characterization workflow to verify whether the defect has a ZPL or not.

In 4H-SiC, there is a systematic ZPL shift of 0.2 eV calculated with PBE functional compared with experimental measurements.
The same shift is expected for other SiC polytypes.
In diamond, there is a slightly larger shift (about 0.3 eV) for the NV-center~\cite{lofgren2018charged}.
In general, one would have to test which shift holds for other selected semiconductor hosts.
More data, both for different defects and host materials, are needed before any accurate conclusion can be made about the systematic shift of the ZPL using the PBE functional.
This shift is for transitions between defect states.
However, for a transition between a defect state and the conduction or valence band edge, the shift can be larger or smaller depending on the defect state position in the band gap.
This is related to the band gap error produced by the PBE functional.
It is well known the bandgap is smaller with the PBE functional than, for example, the HSE functional.
The transitions between defect states are only shifted 0.2 eV but transitions involving the conduction or valence band edges may shift more since there is about a 1 eV difference for the band gap between the PBE functional and experiment.
On this note, it is possible that defect states, which should be in the band gap, could end up either in the conduction or valence band.
In this case, it is difficult to find them since they start mixing with the delocalized states.
However, the workflow can provide some insight.
If a band in the conduction or valence band goes above the threshold for a local band, they are marked as potential defect states.
They are not used in the workflow since occupying and unoccupying them is tricky due to mixing with delocalized bands, but they are marked for future reference.
To verify if these states truly are defect states, one would need to use higher-order methods, like GW or HSE.

The full characterization workflow is thorough, accurate, and only discards runs when they do not converge.
One potential limitation, which is present in both workflows, may be that the neutral charge state is calculated and analyzed first, then the different charge states are calculated based on neutral defect states.
This approach works fine for the silicon vacancy, where only one band seems to move in and out of the band gap depending on the charge state.
However, if the neutral charge state does not have any local state but other charge states do, the workflow would miss these.
This design is chosen to minimize the number of calculations and assumes that the defect bands will not move significantly.
A possible update to the full characterization workflow includes adding the FNV correction and handling potential computational challenges connected with it.
At present, the PBE functional is a necessary choice for carrying out a large number of calculations.
If more accurate calculations are required, higher-order methods can be run by hand or, better yet, constructing additional workflows that use the data present in the database as a starting point.
For example, running the HSE functional on top of the PBE results is an efficient procedure~\cite{methodology_paper}.
Overall, ADAQ provides a strong starting point in the search for novel systems based on defects in semiconductors~\cite{davidsson2021color}.

\section{Conclusion}
\label{sec:conclusion}

We have developed ADAQ which consists of automatic workflows for high-throughput calculations of magneto-optical properties of point defects and their complexes in semiconductors. 
To handle the vast number of possible defects, the following strategy for using ADAQ is proposed.
First, the screening workflow should be employed to consider a large number of defects ($\sim$10 000).
This workflow provides the ZPL and TDM for one potentially interesting transition per defect configuration.
Next, to identify the different configurations, all transitions would be needed.
The full characterization workflow calculates the converged ZPL.
Here, one gets all the single excitations between defect states as well as excitations from the top of the valence band and the bottom of the conduction band for five different charge states of the defect.
In addition to the ZPL, other magneto-optical properties, such as ZFS and hyperfine coupling parameters, are calculated.
With all the calculated properties, accurate identifications of point defects can be made.
The capability of ADAQ is demonstrated on the silicon vacancy in 4H-SiC.
Assuming that this point defect had not already been identified, an accurate identification would have been made with ADAQ.
This demonstrates the potential of the developed collection of workflows for future identification of unknown point defect clusters in any wide band gap semiconductor.
After a potentially interesting candidate has been found, additional manual state-of-the-art calculations can be carried out to better understand the physics of the new point defect.

\appendix
\section{Defect generation}
\label{appendix:defect_gen}

\begin{figure}[h!]
   \includegraphics[width=0.99\textwidth]{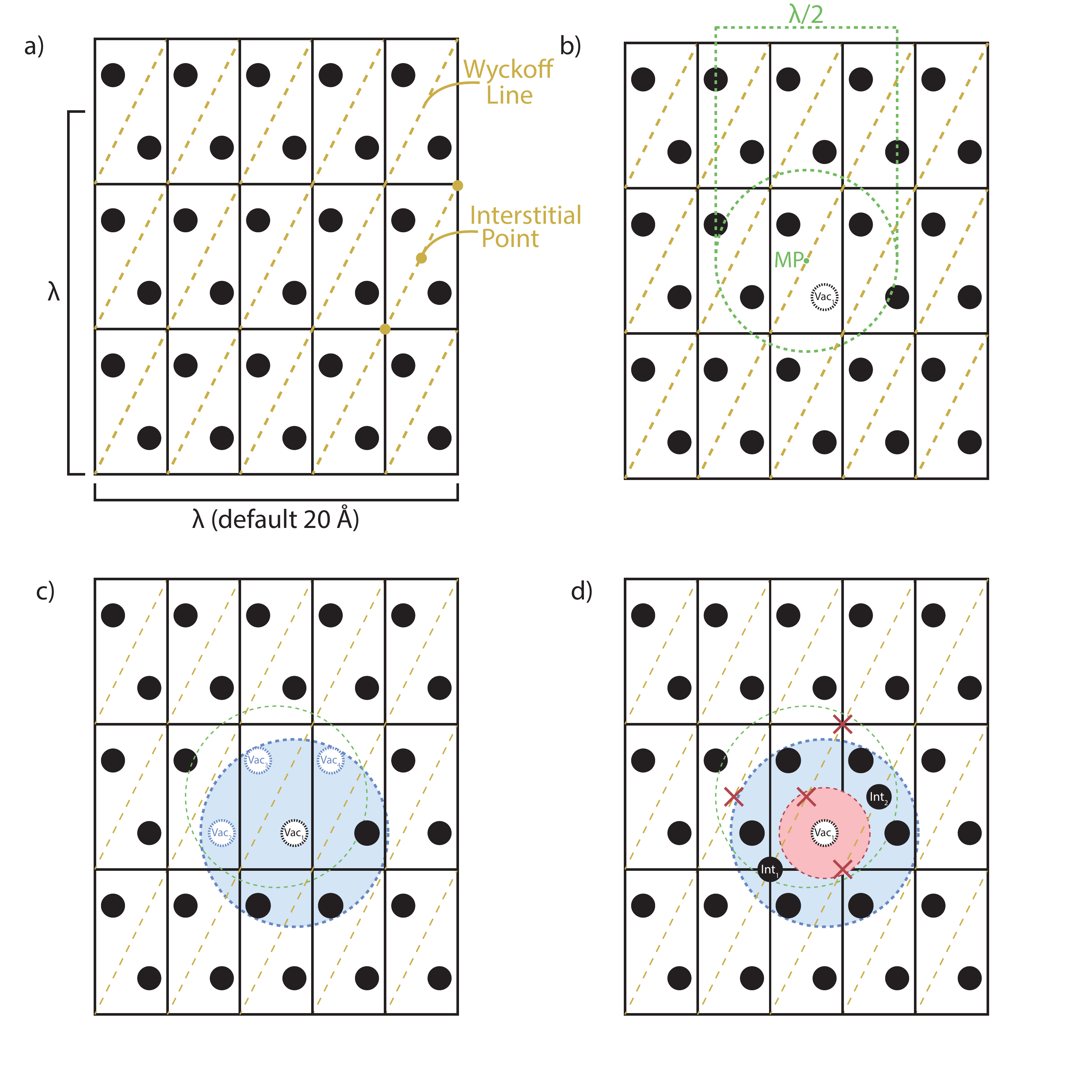}
	\caption{Schematic picture of the defect generation steps. a) shows how the unit cell is copied to be at least $\lambda$ wide, and the interstitial positions are found along a Wyckoff line. b) shows how a single vacancy is generated closest to the midpoint (MP) and inside a distance of half $\lambda$. When the next vacancy is added, only atoms inside both the blue and green spheres are considered, as seen in c). When adding interstitials, two positions are excluded due to being too close to the vacancy, see the red crosses inside the red sphere in d). The two positions outside the blue sphere are also excluded due to being too far away, see the other two red crosses. Thus, only two interstitial positions remain for this defect cluster.}
	\label{fig:def_gen}
\end{figure}

This appendix shows how the defect supercells are generated. 
Figure~\ref{fig:def_gen} shows a schematic picture of the defect generation process.
First, the unit cell is analyzed for symmetry, and non-equivalent atom positions are found.
The interstitial locations are found by a combination of Wyckoff positions and Voronoi tessellation, similar to Ref.~\onlinecite{goyal2017computational}.
The Voronoi locations, found using Voro++~\cite{voro,voro2,voro3}, are mapped to Wyckoff positions, and the interstitial locations are then added to the unit cell in symmetric order (highest symmetry first).
For all interstitial locations, the symmetric copies in the unit cell are also saved.
Any new interstitial location must be farther away from the already found interstitial locations with a minimal interstitial-interstitial distance (default: 0.5 \AA).
After all the interstitial locations have been found, the supercell is constructed by copying the unit cell in x, y, and z-direction so each of the lattice vectors meets at least the input supercell size criterion $\lambda$ (default: 20 Å), see Figure~\ref{fig:def_gen}~a).
For 4H-SiC, this means copying 6, 6, 2 times for a total of 576 atoms.
The defects are generated as close to the midpoint of the cell as possible and inside a sphere with half $\lambda$ as diameter (in this case: 10 Å).
For single vacancies and substitutions, one atom from each of the non-equivalent positions is removed or replaced.
Figure~\ref{fig:def_gen}~b) shows a single vacancy being generated inside the sphere with half $\lambda$ as diameter.
The single interstitials are placed at the predefined positions.

To create point defect clusters, a recursive formula generates combinations of vacancies, substitutions, and interstitials.
Here, two pairwise defect-defect distances are introduced.
First, the largest defect-defect distance between defects is 3.5 Å, which would roughly correlate to the second nearest neighbors in 4H-SiC, to avoid defect clusters where the defects are too far from each other and do not interact.
As seen in Figure~\ref{fig:def_gen}~c), where the three possible second vacancy positions are available.
Only atoms inside both the blue sphere and the green sphere are potential candidates.
Second, the smallest defect-defect distance between a vacancy and an interstitial (1 Å) is introduced to avoid that the interstitial relaxes into the vacancy position, thus creating a substitution.
Figure~\ref{fig:def_gen}~d) shows the potential interstitial positions.
If the interstitial positions are too close or too far away from the vacancy, they are excluded, leaving two potential interstitial positions for a defect cluster consisting of a vacancy and an interstitial.

For these potential positions, there is a possibility that the defect might have been created before or a symmetric equivalent defect cluster exists.
For example, in Figure~\ref{fig:def_gen}~d), there are two possible interstitial positions but these are symmetric copies of each other; hence only one is needed.
To keep track of the generated defects and avoid duplicates, a unique description of the defect clusters is introduced.
The nomenclature for the kind of defect is as follows: defect type+(layer)+interstitial information regarding its location.
Here are some examples for point defects in 4H-SiC: Vac\_Si(h) for a silicon vacancy in the hexagonal layer, C\_Si(k) for a substitution of a silicon with a carbon in the cubic layer, and Int\_Si(h)\_v:p\_w:li\_(0, 0, 3/32) for a silicon interstitial in the hexagonal layer found with a Voronoi point (v:p), on a Wyckoff line (w:li), at the relative coordinates of the unit cell (0, 0, 3/32).
The Voronoi notations denote a point (p), a edge midpoint (e), or a face center point (f); and the Wyckoff notations denote the different symmetry positions like a point (p), a line (li), a plane (pl), or free space (sp).
The relative coordinates are sorted from x to z and rounded to the nearest fraction with 100 as the largest denominator.
In the example above, the interstitial is on the Wyckoff line (x, x, z).
For each defect, a matrix is constructed with the kind of defect on the diagonal and the squared distances between the defects on the off-diagonal.
A point defect cluster like a Frankel pair, a defect consisting of a vacancy and a nearby interstitial, separated by 1.5 Å would have the following matrix:
\begin{equation}
\begin{pmatrix}
\mathrm{Int\_Si(h)\_v:p\_w:li\_(0, 0, 3/32)} & 2.25 \\
2.25 & \mathrm{Vac\_Si(h)} 
\end{pmatrix}
\end{equation}
These symmetric matrices are sorted and saved during the generation.
If the same matrix is found, the present defect cluster is neglected.
This makes sure that only one of the two defect clusters in Figure~\ref{fig:def_gen}~d) is created and stored in the database.
Also, a unique hash index is calculated from these matrices to keep track of the defect in the database.

The defects can also be generated with extrinsic elements.
Here, they can be doped with elements from H through Bi except for the noble gases (He, Ne, Ar, Kr).
Above Bi in the periodic table, elements become increasingly radioactive and thus neglected as dopants.
Since the noble gases do not form any compound, except Xe, they are also neglected as dopants.
One should not expect that the elements with d and f electrons will produce accurate results with the present level of theory, but the defect clusters can be generated never the less.
Depending on the charge state of the defect, d or f orbitals might be unoccupied, and some of them may be interesting to look at.

\section{Screening workflow}
\label{appendix:screen}

\begin{figure}[h!]
   \includegraphics[width=\textwidth]{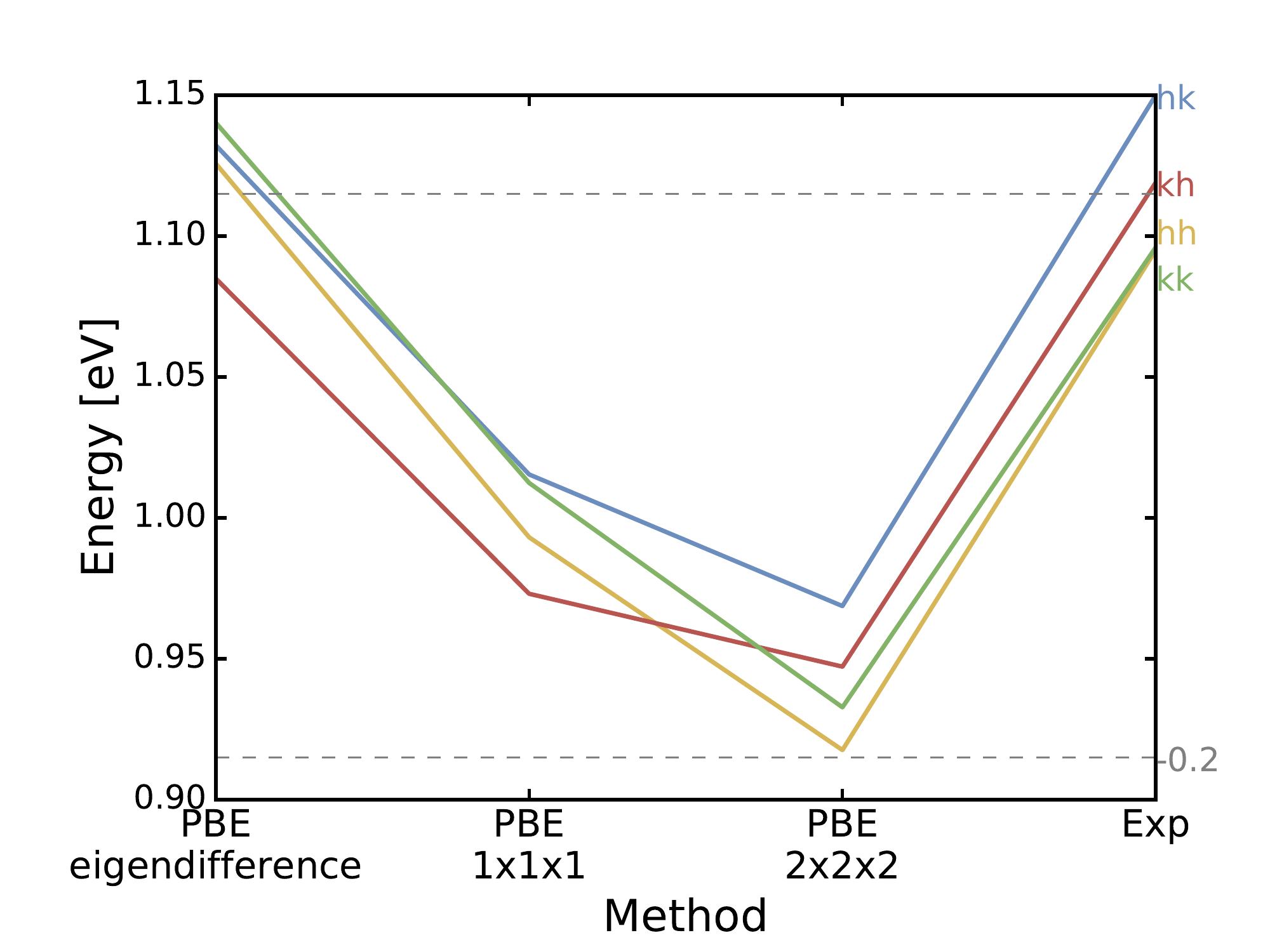}
	\caption{The four configurations of divacancy in 4H-SiC with different convergence settings. Calculated for 576 atom supercell with PBE functional. From right are experimental values, $2 \times 2 \times 2$ k-point set, $\Gamma$ point data taken from Ref.~\onlinecite{methodology_paper}. Furthest to the left is the eigenvalue difference from the screening workflow.}
	\label{fig:comparison} 
\end{figure}

In this appendix, a scaled-down version of the full characterization workflow is presented to screen defects. 
Figure~\ref{fig:comparison} shows different convergence settings used for the ZPL calculations for the divacancy in 4H-SiC.
The full characterization workflow runs with the PBE $2 \times 2 \times 2$ settings.
The order of the configurations is the same as in experiment, and the absolute values are systematically underestimated by about 0.2 eV~\cite{methodology_paper}.
Reducing the k-points to only $\Gamma$-point increases the absolute values but the different configurations can no longer be correctly identified, as shown at PBE $1 \times 1 \times 1$.
This method still requires both ground and excited state calculations to get these values.
However, looking only at the difference between eigenvalues in the ground state, one can estimate the ZPL energy without calculating the excited state.
This difference is marked as PBE eigendifference in the plot where the relaxation of the excited state is neglected, and the energy is surprisingly close to experimental values, at least for the divacancy.
This agreement with experiment is a fortunate error cancellation that the converged PBE values differ by 0.2 eV, and reducing k-point as well as neglecting the relaxation of the excited state makes up this difference.
One cannot assume that this will be the case for all possible defects.
Hence, the eigendifferences are calculated first.
If a eigendifference is larger than 0.4 eV, the excited state with $\Gamma$ point is calculated.
This approach is good enough to determine if a defect is bright or dark and identifies a range that contains the ZPL.

\begin{figure*}[h!]
   \includegraphics[width=\textwidth]{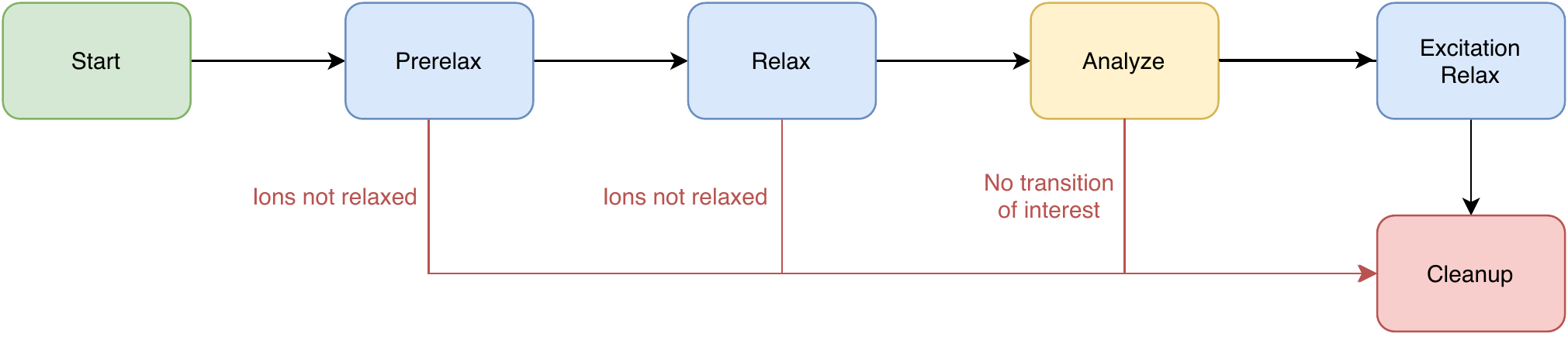}
	\caption{Flowchart of the screening workflow that relaxes and analyzes a defect cell. Green and red boxes show start and end. The blue boxes show VASP calculations, and the yellow box shows the analyze step.}
	\label{fig:screen} 
\end{figure*}

Figure~\ref{fig:screen} shows an overview of the reduced workflow used for screening.
The changes from the ground state workflow are presented in this appendix.
The k-point sampling is reduced from $2 \times 2 \times 2$ to only $\Gamma$-point throughout this workflow.
Hence, the gamma compiled VASP version is used.
The number of charge states is reduced to neutral, plus, and minus; alternative spins are still included.
The final relax step is removed.
The changed VASP settings for all the runs include Fermi smearing, and the FFT grid is set to 3/2 of the largest wave vector.
The following steps have been changed from the ground state workflow, Sec.~\ref{sec:ground}.

\subsection{Prerelax}

Same as in the ground state workflow but now runs with gamma compiled version.

\subsection{Relax}

The k-point set is reduced to $\Gamma$-point only and also runs with gamma compiled version.

\subsection{Analyze}

The same analysis as in the ground state is performed, even though most of the output is not used.
When the defect states have been found, the differences between the occupied and unoccupied eigenvalues are calculated, and the lowest difference is estimated to be the ZPL.
This works as long as there are no singly occupied, almost degenerate, state present for the defect.
To prevent calculating excitations between almost degenerate states, the threshold limit of 0.4 eV is used to exclude any transition below this value.
The smallest transition above this limit is called the transition of interest.
If no such transition exists, then the workflow proceeds to the cleanup step.

\subsection{Excited relax}

This step calculates the transition of interest found in the analyze step with the same settings as in the relax step above.

\subsection{Cleanup}

If everything has finished correctly, then the cleanup removes the WAVECARs as discussed in Sec.~\ref{sec:clean}.
If the defect does not converge because of a particular charge state is unstable or the starting geometry is too far from a stable position, the job is stopped and cleaned up without analyzing it.
These are marked as not converged run and still stored in the database.

\subsection{Database}

Similar data is saved to the database for the screening workflow as for the full characterization workflow.
Additional stored data include the ZPL estimate, TDM, and partial density difference.
The ZPL estimate is taken from the lowest eigenvalue difference.
The TDM and the partial density difference are calculated for the ground state for all local defect states.
The ion relaxation from the input geometry is estimated to show how much the defect relaxed.
If a transition of interest has been calculated, the total energy difference, TDM, and partial density difference between the excited and ground state are also saved to the database.

\section*{Availability}

The data presented in this paper is available via figshare~\cite{figshare}.
For availability of the ADAQ workflow software, see \url{https://httk.org/adaq/}.

\section*{Acknowledgements}

This work was financially supported by the Knut and Alice Wallenberg Foundation through WBSQD2 project (Grant No. 2018.0071).
Support from the Swedish Government Strategic Research Areas in Materials Science on Functional Materials at Linköping University (Faculty Grant SFO-Mat-LiU No. 2009-00971) and the Swedish e-Science Centre (SeRC) are gratefully acknowledged.
VI acknowledges support from the MTA Premium Postdoctoral Research Program.
RA acknowledges support from the Swedish Research Council (VR) Grant No. 2016-04810.
The computations were enabled by resources provided by the Swedish National Infrastructure for Computing (SNIC) at NSC partially funded by the Swedish Research Council through grant agreement no. 2018-05973.

\section*{Acronyms}
\label{sec:acro}
\begin{acronym}[TDMA]
\acro{ADAQ}{Automatic Defect Analysis and Qualification}

\acro{ZPL}{Zero Phonon Line}
\acro{ZFS}{Zero Field Splitting}
\acro{TDM}{Transition Dipole Moment}

\acro{VBM}{Valence Band Maximum}
\acro{CBm}{Conduction Band minimum}

\acro{MP}{Makov-Payne charge correction~\cite{makov1995periodic}}
\acro{LZ}{Lany-Zunger charge correction~\cite{LanyZunger08}}
\acro{FNV}{Freysoldt–Neugebauer–Van de Walle charge correction~\cite{Freysoldt}}

\acro{httk}{The High-Throughput Toolkit~\cite{httk,Armiento2020}}
\acro{VASP}{Vienna Ab initio Simulation Package~\cite{VASP,VASP2}}

\acro{DFT}{Density Functional Theory (DFT)~\cite{Hohenberg64,Kohn65}}
\acro{PBE}{The semi-local exchange-correlation functional of Perdew, Burke, and Ernzerhof~\cite{PBE}} 
\acro{HSE}{The hybrid exchange-correlation functional of Heyd, Scuseria, and Ernzerhof (HSE06)~\cite{HSE03,HSE06}}

\acro{IPR}{Inverse Participation Ratio~\cite{kramer1993localization} -- a measure of localization}
\end{acronym}

\bibliographystyle{apsrev4-1}
\bibliography{references}

\end{document}